\documentclass[preprint,authoryear,12pt]{elsarticle}

\usepackage{graphicx,natbib,amssymb,float,fancyhdr,lastpage}
\usepackage{mathrsfs}
\usepackage{amssymb}
\usepackage{amsmath}
\usepackage{tikz}
\usepackage{color}
\usepackage{paralist}

\usepackage[normalem]{ulem}

\usetikzlibrary{decorations.markings}
\usetikzlibrary{decorations.pathmorphing}
\usetikzlibrary{fit}
\usetikzlibrary{calc,arrows}
\usepgflibrary{plotmarks}

\newcommand{\bg}[1]{\mbox{\boldmath $#1$}}

\newcommand{\tcr}{ \textcolor{black}}

\newcommand{\tr}{\mathrm{tr}\,}

\newcommand{\chain}{\mathrm{chain}}

\newcommand{\M}{\mathrm{max}}

\newcommand{\T}{\mathrm{T}}

\newcommand{\tanhs}{\mathrm{\tanh\,}}

\journal{}

\begin{document}

\begin{frontmatter}
\title{A cyclic stress softening model for the Mullins effect}
\author{S. R. Rickaby}
\ead{stephen.r.rickaby@gmail.com}
\author{N. H. Scott}
\ead{n.scott@uea.ac.uk}
\address{School of Mathematics, University of East Anglia, Norwich Research Park, Norwich NR4 7TJ, UK}

\date{\today}

\begin{abstract}
{In this paper   the inelastic features of stress relaxation, hysteresis and residual strain are combined with the Arruda-Boyce eight-chain model of elasticity, in order 
to develop a model that is capable of describing the Mullins effect for cyclic stress-softening of an  incompressible hyperelastic material, in particular a carbon-filled rubber vulcanizate. We have been unable to identify in the literature any other  model that takes into consideration all the above inelastic features of the cyclic stress-softening of carbon-filled rubber.   Our model compares favourably   with   experimental data and gives a good description of  stress-softening, hysteresis, stress relaxation, residual strain and creep of residual strain.}
\end{abstract}


\end{frontmatter}

\section{Introduction}

\thispagestyle{fancy} \lhead{\emph{Int. J. Solids Structures}  {\bf 50} (2013) 111--120 \hfill
doi:10.1016/j.ijsolstr.2012.09.006}
\cfoot{}

When a rubber specimen is loaded from the virgin state and then unloaded back to this original state, the subsequent load required to produce the same deformation is smaller than that required during primary loading. This phenomenon is known as stress-softening and may be thought of as a decay of elastic stiffness. Stress-softening is particularly evident in specimens of filled rubber vulcanizates.  The vulcanizing of rubber is an irreversible process in which the chemical structure of the rubber is changed in order to improve its elasticity and strength.    During the vulcanization of rubber cross-links are introduced chemically linking the polymer chains together.

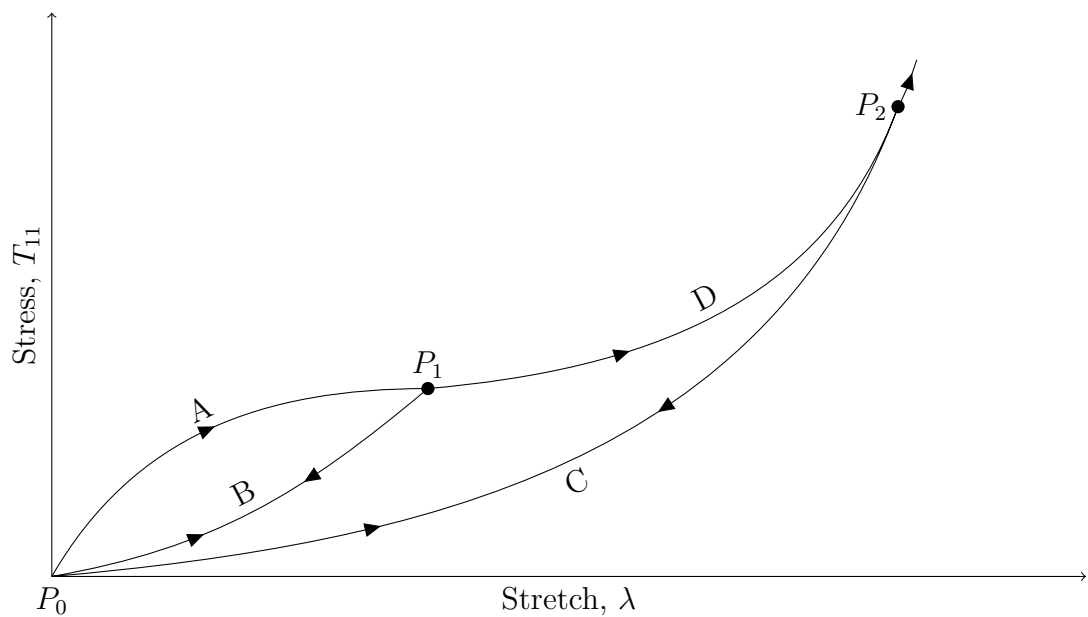
\begin{figure}
\centering
\begin{tikzpicture}[scale=1.25, decoration={
markings,
mark=at position 3cm with {\arrow[black]{triangle 45};},
mark=at position 8.0cm with {\arrow[black]{triangle 45};},
mark=at position 9.5cm with {\arrowreversed[black]{triangle 45};},
mark=at position 16cm with {\arrow[black]{triangle 45};},
mark=at position 20cm with {\arrowreversed[black]{triangle 45};},
mark=at position 28cm with {\arrow[black]{triangle 45};},
mark=at position 33.5cm with {\arrow[black]{triangle 45};}},
]
\draw[->] (0,0) -- (11,0)node[sloped,below,midway] {Stretch, $\lambda$};
\draw[->] (0,0) -- (0,6) node[sloped,above,midway] {Stress, $T_{11}$};
\fill (4,2) circle (2pt);
\fill (9,5) circle (2pt);
\draw [postaction={decorate}](0,0) to[out=60,in=180] node [sloped,above] {A} (4,2)
(0,0) to[out=10,in=220] node [sloped,above] {B} (4,2)
(0,0) to[out=5,in=250] node [sloped,below] {C} (9,5)
(4,2) to[out=5,in=250] node [sloped,above] {D} (9,5)
(9,5) to[out=60,in=250] node [sloped,above] {} (9.2,5.5);
\coordinate [label=above:{$P_1$}] (A) at (4,2);
\coordinate [label=left:{$P_2$}] (B) at (9,5);
\coordinate [label=below:{$P_0$}] (O) at (0,0);
\end{tikzpicture}
\caption{The idealized behaviour of stress-softening in rubber.}
\label{fig:1y}
\end{figure}

Figure \ref{fig:1y} represents the idealized stress-softening behaviour of a rubber specimen under uniaxial tension. The process starts from an unstressed and unstrained virgin state at point $P_0$ and time $t_0$.  Subsequently,  the stress/strain relation follows path $A$, the primary loading path, until point $P_1$ is reached at a time $t_1$. At this point $P_1$, unloading of the specimen begins immediately and the stress/strain relation of the rubber  follows the new path $B$, which lies below $A$, 
returning to the unstressed and unstrained state at point $P_0$. If the material is then reloaded the stress-strain relation  follows path $B$ again, rather than path $A$, up to point $P_1$. If the rubber is now strained beyond point $P_1$ then path $D$ is activated,  a continuation of the original primary loading path. If subsequent unloading occurs from the point $P_2$, the rubber  retracts along a new path $C$ to the unstressed state at $P_0$. The shape of this second stress-strain cycle differs significantly from the first. If the material is now reloaded the stress-strain behaviour follows the new path $C$, rejoining the  primary loading path at the point $P_2$.   

This stress-softening phenomenon is known as the Mullins effect, named after Mullins who conducted an extensive study  into carbon-filled rubber vulcanizates, see  \cite{mullins1947}. \cite{dianib} have written a  recent review of this effect, detailing specific features associated with stress-softening and providing a pr\'ecis of models developed to represent this effect.

Many authors have modelled the Mullins effect since Mullins, see, for example, \cite{ogden}, \cite{dorfmann}, \cite{dianib} and the references cited therein.  However, most authors  model a simplified version of the Mullins effect, in which the following inelastic features are neglected:
\begin{compactitem}
\item Hysteresis
\item Stress relaxation
\item Residual strain
\item Creep of residual strain
\end{compactitem}
Despite the wealth of research into the Mullins effect over the last six or more decades we have been unable to identify in the literature any other model that has been used to reproduce the Mullins effect for cyclic stress-softening of unrefined experimental data.

\pagestyle{fancy}
\fancyhead{}
\fancyhead[RO,LE]{\thepage}
\fancyhead[LO]{{A cyclic stress softening model for the Mullins effect}}
\fancyhead[RE]{N. H. Scott}

\tcr{For a typical carbon filled rubber vulcanizate it is evident from the experimental data presented in Figures \ref{fig:14y}, \ref{fig:15y} and \ref{fig:16y} that cyclically stretched carbon filled rubber vulcanizates undergo hysteresis, stress-relaxation, residual strain and creep of residual strain. 
 Here, the unloading and reloading curves do not follow the same path and are positioned away from the origin with the successive relaxation paths being situated each below the previous one.}

In this paper we derive an isotropic constitutive model to represent the Mullins effect for cyclic stress-softening under uniaxial tension.
We explore  the notion that in order to develop a more realistic model of stress softening  the above inelastic features must  be included. However, not all softening features may be relevant for a particular application, and so in order to develop a functional model we require that specific parameters could be set to zero to exclude any particular inelastic feature above, and still maintain the integrity of the model.  

 The paper is constructed as follows.   In Section \ref{sec:elasticity} we describe the purely elastic response to the initial uniaxial tension in an incompressible isotropic non-linear elastic solid.  We employ the elastic model of \cite{arruda} but any other model of incompressible isotropic elasticity could be employed in its stead.
 In Section \ref{sec:softening} we define the softening function which forms the basis of our new model for softening in uniaxial tension.    Also in Section \ref{sec:softening} we remodel the softening function of \cite{dorfmann2003}  to control the rate of softening and include the effects of hysteresis by defining separate softening functions for  unloading and reloading.  This is also a model for hysteresis, see  \cite{johnsonb}.    Section \ref{sec:relaxation} models the effects of stress relaxation  following the development of \cite{bernstein} and \cite{lockett} and applies the model to cyclic stress relaxation. Section \ref{sec:residual} presents a discussion of residual strain that is motivated by the work of \cite{bergstrom} who develop  a residual strain model by regarding it as a form of creep.  In Section \ref{sec:creepresidual} we discuss the creep of residual strain and develop the \cite{bergstrom} model to account for it.    In Section \ref{sec:constitutive} the models developed in Sections \ref{sec:elasticity}--\ref{sec:creepresidual} are combined to obtain our new  model for  stress-softening of an incompressible isotropic elastic material in uniaxial tension which incorporates all the inelastic effects discussed in this paper. Finally, in Section \ref{sec:numerical}  a graphical presentation of the model is provided and a comparison made between the  model predictions  and  experimental data and conclusions are discussed in Section \ref{sec:conclusion}.
 
 Preliminary results of the model were presented in \cite{ricscott}. 

\section{Elastic response}
\label{sec:elasticity}

In the reference configuration  a material particle is  located at position $\textbf{X}$ at time $t_0$ with Cartesian components $X_1,X_2,X_3$.   After  deformation the same particle is located  at the position $\bg{x}(\textbf{X}, t)$,  at time 
$t$, with Cartesian components $x_1,x_2,x_3$.   The deformation gradient is defined by
\begin{equation*}
F_{iA}(\textbf{X},t)=\frac{\partial x_i(\textbf{X},t)}{\partial X_A},\quad{\rm or\;\;simply,}\quad
\textbf{F}(\textbf{X},t)=\frac{\partial \bg{x}(\textbf{X},t)}{\partial \textbf{X}}.
\end{equation*}
An isochoric uniaxial strain is taken in the form
\[
x_1=\lambda X_1, \quad x_2=\lambda^{-\frac{1}{2}} X_2, \quad x_3=\lambda^{-\frac{1}{2}} X_3,
\]
with $\lambda$ denoting the uniaxial stretch.
The left Cauchy-Green strain tensor $\textbf{B}=\textbf{F}\textbf{F}^\T$  is given by
\begin{displaymath}
\textbf{B}= \left( \begin{array}{ccc}	
{\lambda^2}& 0 &0\\
0 &{\lambda^{-1}}& 0\\
0& 0 &{\lambda^{-1}} \end{array}\right),
\end{displaymath}
with principal invariants
\begin{equation}
 I_1 = \tr {\bf B} = \lambda^2+2\lambda^{-1},\qquad I_2 = I_3 \,\tr {\bf B}^{-1} = \lambda^{-2}+2\lambda,
\qquad I_3 = 1, 
\label{eq:1z}
\end{equation}
the last following from isochoricity.

An incompressible isotropic hyperelastic material possesses a strain energy function $W(I_1, I_2)$ 
per unit volume in terms of which the Cauchy stress is given by
\begin{equation}
\textbf{T}^{\mathscr{E}}(\lambda) = \,  -p \textbf{I}+2\left[\frac{\partial{W}}{\partial I_1}+I_1\frac{\partial{W}}{\partial I_2}\right]
\textbf{B}-2\frac{\partial{W}}{\partial I_2}\textbf{B}^2,
\label{eq:2z}
\end{equation}
with $p $ an arbitrary pressure and $\bf I$ the unit tensor. 
We are concerned here only with uniaxial tension in the 1-direction and so may fix the value of $p$ by the requirement
\[  T_{22}^{\mathscr{E}}(\lambda) =  T_{33}^{\mathscr{E}}(\lambda)  =0.  \]
Using this value of $p$ in equation (\ref{eq:2z}) then gives the only non-zero component of stress to be
the uniaxial tension
\begin{equation}
 T_{11}^{\mathscr{E}}(\lambda) =  2(\lambda^2 - \lambda^{-1})
 \left[\frac{\partial{W}}{\partial I_1}+  \lambda^{-1}\frac{\partial{W}}{\partial I_2}\right],
\label{eq:3z}
\end{equation}
in which equation (\ref{eq:1z})$_1$ has been used.

Equation (\ref{eq:3z}) then gives the uniaxial stress $T_{11}^{\mathscr{E}}(\lambda) $ on the primary loading path $A$ of Figure \ref{fig:1y}.
Denote by $\lambda_\M$ the value of the uniaxial stretch $\lambda$ at the point $P_1$ on Figure~\ref{fig:1y}.  Along path $A$ of  Figure \ref{fig:1y} we therefore have $1\leq\lambda\leq\lambda_\M$.  

\subsection{The Arruda-Boyce eight-chain model}
We now specialize to a particular model of incompressible non-linear isotropic elasticity both for definiteness and for  comparison with experimental data in Section \ref{sec:experimentaldata}. We have selected the Arruda-Boyce eight-chain model as it requires only the measurement of two physical parameters.   Furthermore, it has been shown to provide a good fit to experimental data, see \cite{zuniga}.    Rubber is composed of polymer chains, with each polymer chain being made-up of single links called monomers.
This motivates the eight-chain model of \cite{arruda} based upon the structure of a cube with eight polymer chains joining the corners of the cube to the central point.  In the undeformed cube each such chain has the same length.  If the deformation of the elastic material is such that the cube is deformed into a cuboid it remains true that all the chains have equal  length, though in general a length different from that of the chain length associated with the original cube.
Using this fact and arguments based on statistical mechanics \cite{arruda} showed that the strain energy must take the form
\begin{equation}
W=\mu {N}\left\{\left[\sqrt{\frac{I_1}{3{{N}}}}\right] \mathscr{L}^{-1}\left(\sqrt{\frac{I_1}{3{{N}}}}\right)+\log \left\{\frac{\mathscr{L}^{-1}\left(\sqrt{\frac{I_1}{3{{N}}}}\right)}{\sinh\left(\mathscr{L}^{-1}\left(\sqrt{\frac{I_1}{3{{N}}}}\right)\right)}\right\} \right\}-h_0,
\label{eq:4z}
\end{equation}
where $\mu$ is the positive ground state shear modulus,  ${N}$ is the number of links forming a single polymer chain and $h_0$ is a constant such that the strain energy vanishes in the  undeformed state.
$\mathscr{L}^{-1}(x)$ is the inverse function of the Langevin function
\[
\mathscr{L}(x)=\coth x-\frac{1}{x}.
\]

Upon substituting for $W$ from equation (\ref{eq:4z}) into equation (\ref{eq:2z}) we obtain the stress in the Arruda-Boyce model:
\begin{align}
\textbf{T}^{\mathscr{E}}(\lambda) =& \, -p \textbf{I}+ \mu \sqrt{\frac{N}{3I_1}}\mathscr{L}^{-1}\left(\sqrt{\frac{I_1}{3{N}}}\right)\textbf{B},
\label{eq:5z}
\end{align}
and equation (\ref{eq:3z}) for the single uniaxial stress reduces to 
\begin{equation}
 T_{11}^{\mathscr{E}}(\lambda) =  2\mu(\lambda^2 - \lambda^{-1})
\sqrt{\frac{N}{3I_1}}\mathscr{L}^{-1}\left(\sqrt{\frac{I_1}{3{N}}}\right).
\label{eq:6z}
\end{equation}

\section{Softening function and hysteresis}
\label{sec:softening}
\cite{zuniga} defined the Cauchy stress $\textbf{T}$ in the  unloading and reloading of the  material to be a product of the Cauchy stress $\textbf{T}^{\mathscr{E}}(\lambda)$ in an isotropic elastic parent material, as in
Section \ref{sec:elasticity},  and a softening function $\zeta(m)$ that depends on the current value of the magnitude of the strain $m$, where $m\equiv \sqrt{\textbf{B}\cdot\textbf{B}}$.  However, in the present case of uniaxial tension it is more convenient to take as measure of strain magnitude simply the uniaxial strain $\lambda$, so that the softening function takes the form $\zeta(\lambda)$.  We then have
\begin{equation}
\textbf{T}=\zeta(\lambda)\textbf{T}^{\mathscr{E}}(\lambda)
\label{eq:7z}
\end{equation}
along the unloading path $B$ of Figure \ref{fig:1y}.   At point $P_1$ the stresses on paths $A$ and $B$ must be equal and so equation (\ref{eq:7z}) requires $\zeta(\lambda_\M) = 1$.  Path $B$ must be below path $A$ and yet give positive stresses and so the function $\zeta(\lambda)$ must satisfy
\begin{equation}
 0 < \zeta(\lambda)\leq 1\quad\mbox{for}\quad 1\leq\lambda\leq\lambda_\M, 
 \label{eq:8z}
 \end{equation}
with equality only for $\lambda=\lambda_\M$.

\cite{dorfmann2003,dorfmann}, employing a theory of pseudo-elasticity, effectively proposed as softening function 
\begin{equation}
\zeta(\lambda)=1-\frac{1}{r}\tanhs\left(\frac{W_\M-W}{\mu b}\right),
\label{eq:9z}
\end{equation}
in which $b$ and $r$ are positive dimensionless material parameters.  Choosing $r\geq 1$ ensures that 
 $\zeta(\lambda)>0$ for all choices of $b$.    As before, $\mu$ is the  ground state shear modulus.    $W_\M$ is the value which the strain energy $W$  takes at the point $P_1$.  For stability it must be that the strain energy $W$ is a monotonically increasing function of the uniaxial stretch $\lambda$.  Since $I_1$, defined by equation (\ref{eq:1z})$_1$, is a monotonically increasing function of $\lambda$ in $\lambda\geq 1$, it follows that, for stability under uniaxial tension, $W$ must be a monotonically increasing function of $I_1$.  Therefore $0\leq W\leq W_\M$  on the primary loading path $A$.  It follows that $\zeta(\lambda)$ defined by (\ref{eq:9z}) satisfies the inequalities (\ref{eq:8z}) and is monotonically increasing on $\lambda\geq 1$.
It can be verified that the Arruda-Boyce strain energy function (\ref{eq:4z}) is monotonically increasing on $\lambda\geq 1$.

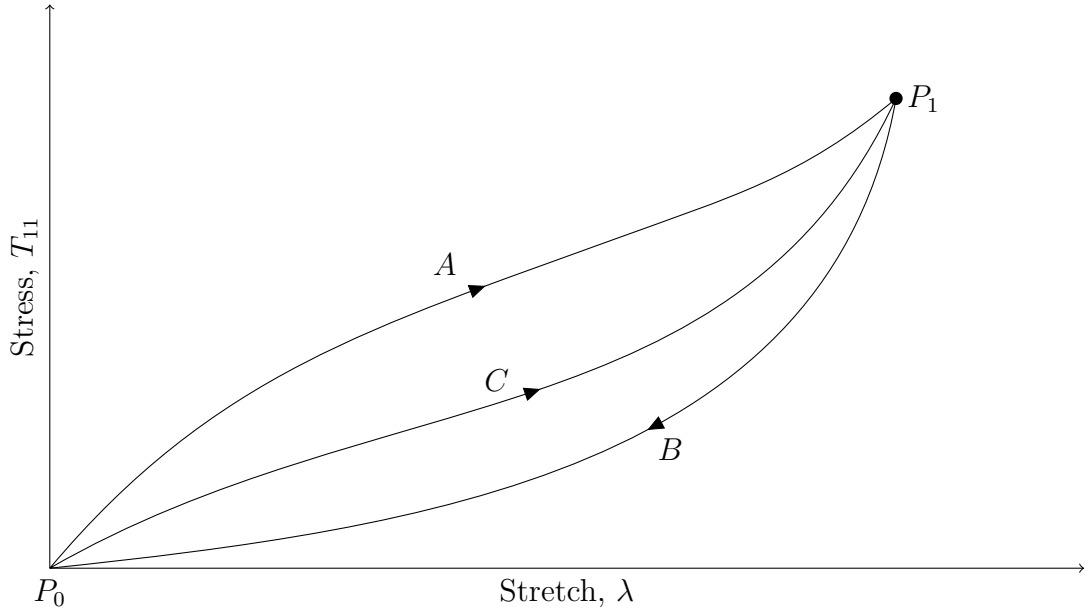
\begin{figure}[ht]
\centering
\begin{tikzpicture}[scale=1.25, decoration={
markings,
mark=at position 7cm with {\arrow[black]{triangle 45};},
mark=at position 20cm with {\arrow[black]{triangle 45};},
mark=at position 34.5cm with {\arrowreversed[black]{triangle 45};},}
]
\draw[->] (0,0) -- (11,0)node[sloped,below,midway] {Stretch, $\lambda$};
\draw[->] (0,0) -- (0,6) node[sloped,above,midway] {Stress, $T_{11}$};
\fill (9,5) circle (2pt);
\draw [postaction={decorate}]
(0,0) to[out=50,in=200] node [sloped,above] {} (6,3.5)
(6,3.5) to[out=20,in=220] node [sloped,above] {} (9,5)
(0,0) to[out=30,in=245] node [sloped,above] {} (9,5)
(0,0) to[out=6,in=260] node [sloped,below] {} (9,5);
\coordinate [label=above:{$A$}] (A) at (4.2,3.0);
\coordinate [label=left:{$C$}] (B) at (5.0,2.0);
\coordinate [label=below:{$B$}] (B) at (6.6	,1.5);
\coordinate [label=right:{$P_1$}] (B) at (9,5);
\coordinate [label=below:{$P_0$}] (O) at (0,0);
\end{tikzpicture}
\caption{Stress-softening including hysteresis.}
\label{fig:3y}
\end{figure}

\cite[Section 3.6]{johnsonb} observed that a typical stress-stretch response, that is obtained experimentally, is as depicted in Figure \ref{fig:3y}.  Initial loading follows the path $A$ up to a point $P_1$ and subsequent unloading follows the path $B$, which lies below $A$. Reloading to the strain at $P_1$  then follows the new path $C$, which lies between $A$ and $B$.  Subsequent unloading from $P_1$ to $P_0$ and reloading from $P_0$ to  $P_1$ follows paths $B$ and $C$, respectively.   The fact that path $C$ lies above path $B$, and does not coincide with it, constitutes the phenomenon of hysteresis.

\subsection{The softening functions for unloading and reloading}
\label{sec:softening1}

We shall model this   unloading and reloading feature by introducing a dimensionless parameter $\vartheta_\omega$ in order to control the rate at which the  hyperbolic tangent term in equation (\ref{eq:9z}) tends to zero as its argument tends to zero. We may then introduce a constant $\vartheta_1$ for  unloading and a different constant $\vartheta_2$ for  reloading.  It turns out that we  need also to give the parameters $r$ and $b$ in equation (\ref{eq:9z}) different values in unloading and reloading.
Therefore we shall replace the softening function of equation  (\ref{eq:9z}) by
\begin{equation}
 \zeta_\omega(\lambda)=1-\frac{1}{r_\omega}\left\{\tanhs\left(\frac{W_\M-W}{\mu b_\omega}\right)\right\}^{1/\vartheta_\omega},
\label{eq:10z}
\end{equation}
where
\[
\omega=\left\{ \begin{array}{clrr}
1 & \textrm{ unloading},\quad \textrm{following path $B$},\\[2mm]
2 & \textrm{ reloading},\quad \textrm{following path $C$}.\\
\end{array}\right.
\]
Choosing the parameters so that
\[ r_2\geq r_1\geq 1,\quad b_2\geq b_1,\quad \vartheta_2\geq \vartheta_1, \]
with not all equalities holding together,
guarantees that path $C$ lies above path $B$, yet remains below path $A$, and that
the inequalities (\ref{eq:8z}) remain in force, taking here the form
\[ 0 <  \zeta_1(\lambda) <  \zeta_2(\lambda) < 1\quad\mbox{for}\quad 1\leq\lambda<\lambda_\M,   \]
with $\zeta_1(\lambda_\M) =  \zeta_2(\lambda_\M) =1$ continuing to hold.

On the primary loading path $A$ of Figure \ref{fig:3y} the elastic stress is given by $\textbf{T}^{\mathscr{E}}(\lambda)$, exactly as on the same path of Figure \ref{fig:1y}.  Upon paths $B$ and $C$ of Figure \ref{fig:3y} the stress is given by modifying (\ref{eq:7z}) to read
\begin{equation}
\textbf{T}=    \zeta_\omega(\lambda)  \textbf{T}^{\mathscr{E}}(\lambda),
\label{eq:11z}
\end{equation}
where $ \zeta_\omega(\lambda)$ is given by equation  (\ref{eq:10z}), with
$\omega=1$ corresponding to the unloading path $B$ and $\omega=2$ to the reloading path $C$.

When fitting this model to experimental data it is observed that as the stretch increases the stress relaxation paths underpredict the stress. By altering the softening parameter $b$, we can alter the curvature of the paths.  This motivates  assigning one softening parameter $b_1$ to the  unloading path and a different softening parameter $b_2$ to the  reloading path.
The inclusion of the additional parameter $\vartheta_\omega$ in the \cite{dorfmann2003,dorfmann} model introduces further control of the shape of the softening function.

Equation (\ref{eq:11z}) constitutes a model for hysteresis because it gives a reloading path $C$ that is different from the unloading path $B$ and lies above it.  It does this by the introduction of a softening function   (\ref{eq:10z}) which is different on each of the paths $B$ and $C$.  This model is capable of representing the Mullins effect over multiple cycles of hysteresis and stress-softening.  This approach has not  previously been considered in the literature. 

\section{Cyclic stress relaxation}
\label{sec:relaxation}
Suppose a material body is deformed in some way by applied stresses and is then held in the same state of deformation over a period of time by  applied stress.
Stress relaxation is said to occur if the stress needed to maintain this fixed deformation decreases over the period of time. 

\begin{figure}[ht]
\centering
\begin{tikzpicture}[scale=1.25, decoration={
markings,
mark=at position 7cm with {\arrow[black]{triangle 45};},
mark=at position 15cm with {\arrowreversed[black]{triangle 45};},
mark=at position 29cm with {\arrow[black]{triangle 45};},}
]
\draw[->] (0,0) -- (11,0)node[sloped,below,midway] {Stretch, $\lambda$};
\draw[->] (0,0) -- (0,6) node[sloped,above,midway] {Stress, $T_{11}$};
\draw [black](0,0) to[out=50,in=200] node [sloped,above] {} (6,3.5);
\draw [black](6,3.5) to[out=20,in=220] node [sloped,above] {} (9,5);
\draw [black] (0,0) to [out=6,in=260] node [sloped,above] {} (9,5);
\draw [black!40] (0,0) to[out=40,in=240] node [sloped,above] {} (9,4.5)
(0,0) to[out=6,in=256.5] node [sloped,below] {} (9,4.5);
\draw [dotted](0,0) to[out=6,in=253] node [sloped,above] {} (9,4.1)
(0,0) to[out=40,in=238] node [sloped,below] {} (9,4.1);
\draw [dashed] (0,0) to[out=6,in=250] node [sloped,above] {} (9,3.85)
(0,0) to[out=40,in=236] node [sloped,below] {} (9,3.85);
\draw [postaction={decorate}][loosely dotted, line width=0.01pt](0,0) to[out=50,in=200] node [sloped,above] {} (6,3.5)
(0,0) to [out=6,in=260] node [sloped,above] {} (9,5)
(0,0) to[out=40,in=240] node [sloped,above] {} (9,4.5)
(0,0) to[out=6,in=256] node [sloped,below] {} (9,4.5);
\coordinate [label=right:{$P_1,\,t_1$}] (B) at (9.1,5);
\coordinate [label=right:{$P_2,\,t_2$}] (B) at (9.1,4.5);
\coordinate [label=right:{$P_3,\,t_3$}] (B) at (9.1,4.1);
\coordinate [label=right:{$P_4,\,t_4$}] (B) at (9.1,3.8);
\coordinate [label=right:{$A$}] (B) at (4.2,3.3);
\coordinate [label=right:{$C$}] (B) at (4.2,2.1);
\coordinate [label=right:{$B$}] (B) at (4.6,1.1);
\coordinate [label=below:{$P_0$}] (O) at (0,0);
\end{tikzpicture}
\caption{Cyclic stress relaxation.}
\label{fig:4y}
\end{figure}
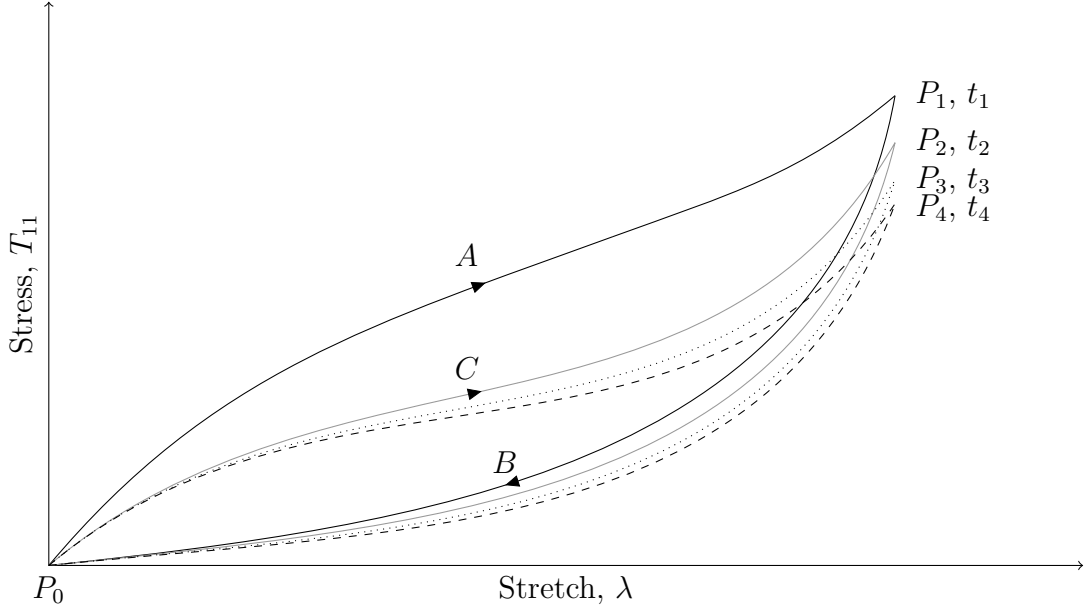

When a carbon filled rubber vulcanizate is cyclically loaded and unloaded to a specified strain, 
\cite[Figure 1]{holt} observed that the successive relaxation paths are situated each below the previous one.
This is illustrated in Figure~\ref{fig:4y}, where the primary loading path $P_0P_1$ (i.e. path $A$) lies above the first reloading path $P_0P_2$ (i.e. path $C$) which, in turn, lies above the second reloading path $P_0P_3$, and so on.  
Similarly, the first unloading path $P_1P_0$ (i.e. path $B$) lies above the second unloading path $P_2P_0$, and so on.
Eventually,  equilibrium reloading and unloading paths are reached.
 \cite{fletcher} found experimentally that for rubber vulcanizates  this viscoelastic  stress relaxation is non-linear.

\cite{dannenberg} proposed that during primary loading of a vulcanized rubber, which is composed of polymer chains, some of the  molecular cross-links or bonds between chains undergo slippage and breaking and that during subsequent stress relaxation some, but not all, of these broken bonds reform and the slippage partially recovers to its original position.  In the present model we assume that the material does not relax during primary loading so that stress-relaxation  occurs only during  the subsequent unloading and reloading phases,  commencing at  time $t_1$,  the time at which primary loading ceased.

Figure \ref{fig:4y} represents a cyclically loaded and unloaded rubber specimen with primary loading occurring along  path $P_0P_1$ up to the point $P_1$ where $\lambda=\lambda_\M$,  which is reached at  time $t_1$.   Stress-relaxation then commences at time $t_1$ and follows the   unloading  path $P_1P_0$  back to the position $P_0$ of zero stress, which is reached at time $t^*_1$.  On the  reloading  path $P_0P_2$ stress-relaxation proceeds until point $P_2$ is reached, at time $t_2$, where once again $\lambda=\lambda_\M$.    Stress-relaxation continues as we follow the grey  unloading path $P_2P_0$ to the position $P_0$  of zero stress, reached at time $t^*_2$. This pattern then continues throughout the unloading and reloading process.  Stress-relaxation may proceed at different rates in  unloading and  reloading, i.e. $t_1^{*}-t_1^{\phantom{*}}$ and $t_2^{\phantom{*}}-t_1^{*}$ may be unequal.

\subsection{The Bernstein, Kearsley and Zapas model}

\cite{bernstein} developed a model, known as the BKZ model, for non-linear stress relaxation,  postulating that if a material is observed at sufficiently low temperatures and for a short time, then it is difficult to distinguish its behaviour from that of an elastic solid. When the material is observed for a sufficiently long time and at a sufficiently high temperature, flow behaviour is more significant. The BKZ model has been found to  represent accurately experimental data for stress-relaxation, see \cite{tanner} and the references therein.

For an incompressible viscoelastic solid,  \cite[pages 114--116] {lockett} derived the following version of the \cite{bernstein} model for the relaxation stress $\textbf{T}^{\mathscr{R}}$:
\begin{equation}
\textbf{T}^{\mathscr{R}}(\lambda, t)=-p \textbf{I} + \Bigg[{A}_0+\frac{1}{2}    \breve{A}_1(t)(I_1-3)-  \breve{A}_2(t)\Bigg]\textbf{B}+  \breve{A}_2(t){\textbf{B}}^{2},\quad\mbox{for}\;\; t>t_1.
\label{eq:12z}
\end{equation}
 For  uniaxial tension we argue  as before to determine the pressure $p$ from the requirement that  $T^{\mathscr{R}}_{22}=T^{\mathscr{R}}_{33}=0$ in equation  
 (\ref{eq:12z}),   and then use this value of $p$ to show that the only non-zero component of stress
 in (\ref{eq:12z})  is   the uniaxial tension
\begin{align}
T^{\mathscr{R}}_{11}(\lambda, t)=&\, (\lambda^2-\lambda^{-1})
\left[{A}_0+\frac{1}{2} \breve{A}_1(t)
(\lambda-1)^2(1+2\lambda^{-1})
+ \breve{A}_2(t)(\lambda^2 -1 + \lambda^{-1} )\right],
\label{eq:13z}
\end{align}
with $T_{11}^{\mathscr{R}}(\lambda, t)$ vanishing for $t\leq t_1$.
In (\ref{eq:12z}), ${A}_0$ is a material constant and  $  \breve{A}_1(t),   \breve{A}_2(t)$ are material functions  which vanish for $t\leq t_1$ and are continuous for all $t$.

As a consequence of the above discussion for cyclic stress-relaxation  the material functions $\breve{A}_1(t)$ and $  \breve{A}_2(t)$  are replaced by
\begin{equation}
A_{1, 2}(t)=\left\{\begin{array}{llll}
   0                                      & \textrm{primary loading}, & t_0^{\phantom{*}}\leq t\leq t_1^{\phantom{*}}, & \textrm{path}\;\; P_0P_1\\[0.5mm]
  \breve{A}_{1, 2}(\phi_1(t-t_1))& \textrm{unloading}, & t_1^{\phantom{*}}\leq t\leq t_1^*, &  \textrm{path}\;\; P_1P_0\\[0.5mm]
  \breve{A}_{1, 2}(\phi_2(t-t_1))& \textrm{reloading}, & t^*_1\leq t\leq t_2^{\phantom{*}}, &  \textrm{path}\;\; P_0P_2\\[0.5mm]
  \breve{A}_{1, 2}(\phi_1(t-t_1))& \textrm{unloading}, & t_2^{\phantom{*}}\leq t\leq t_2^*, &  \textrm{path}\;\; P_2P_0\\[0.5mm]
  \breve{A}_{1, 2}(\phi_2(t-t_1))& \textrm{reloading}, & t^*_2\leq t\leq t_3^{\phantom{*}}, &  \textrm{path}\;\; P_0P_3\\[0.5mm]
 \;\; \dots&\;\; \dots&\;\; \dots&\;\; \dots
\end{array}\right.
\label{eq:14z}
\end{equation}
with $\phi_1$ and $\phi_2$ being continuous functions of time.  Separate functions $\phi_1$ and $\phi_2$ are needed for the unloading and reloading phases, respectively, in order to model better the experimental data in
Section \ref{sec:experimentaldata}.

Employing equation (\ref{eq:14z}), equation (\ref{eq:13z})  becomes,
\begin{equation}
T^{\mathscr{R}}_{11}(\lambda, t)= (\lambda^2-\lambda^{-1})
\left[{A}_0+\frac{1}{2} {A}_1(t)
(\lambda-1)^2(1+2\lambda^{-1})
+{A}_2(t)(\lambda^2 -1 + \lambda^{-1} )\right].
\label{eq:15z}
\end{equation}

\begin{figure}[ht]
\centering
\begin{tikzpicture}[scale=1.25]
\draw[->] (0,6) -- (11,6)node[sloped,below,midway] {} ;
\draw[<-] (0,0) -- (0,6) node[sloped,above,midway] {$A_{1, 2}(t)$};
\coordinate [label=above:{$t$}] (B) at (5.5	,6.5);
\coordinate [label=above:{$t_1^{\phantom{*}}$}] (B) at (1.0	,6.05);
\coordinate [label=above:{$t_2^{\phantom{*}}$}] (B) at (4	,6.05);
\coordinate [label=above:{$t_3^{\phantom{*}}$}] (B) at (7.0	,6.05);
\coordinate [label=above:{$t_1^*$}] (B) at (2.5	,6.05);
\coordinate [label=above:{$t_2^*$}] (B) at (5.5	,6.05);
\coordinate [label=above:{$t_3^*$}] (B) at (8.5	,6.05);
\coordinate [label=above:{$t_0^{\phantom{*}}$}] (O) at (0.0,6.05);
\draw [black]
(1,6) to[out=280,in=180] node [sloped,above] {} (2.5,5.0)
(2.5,5.0) to[out=300,in=180] node [sloped,above] {} (4,4.5)
(4,4.5) to[out=280,in=180] node [sloped,above] {} (5.5,3.5)
(5.5,3.5) to[out=300,in=180] node [sloped,above] {} (7,3)
(7,3) to[out=280,in=180] node [sloped,above] {} (8.5,2.0);
\draw [black][dotted] 
(8.5,2) to[out=300,in=180] node [sloped,above] {} (10,1.5);
\end{tikzpicture}
\caption{Graphical representation of $A_{1, 2}(t)$.}
\label{fig:5y}
\end{figure}
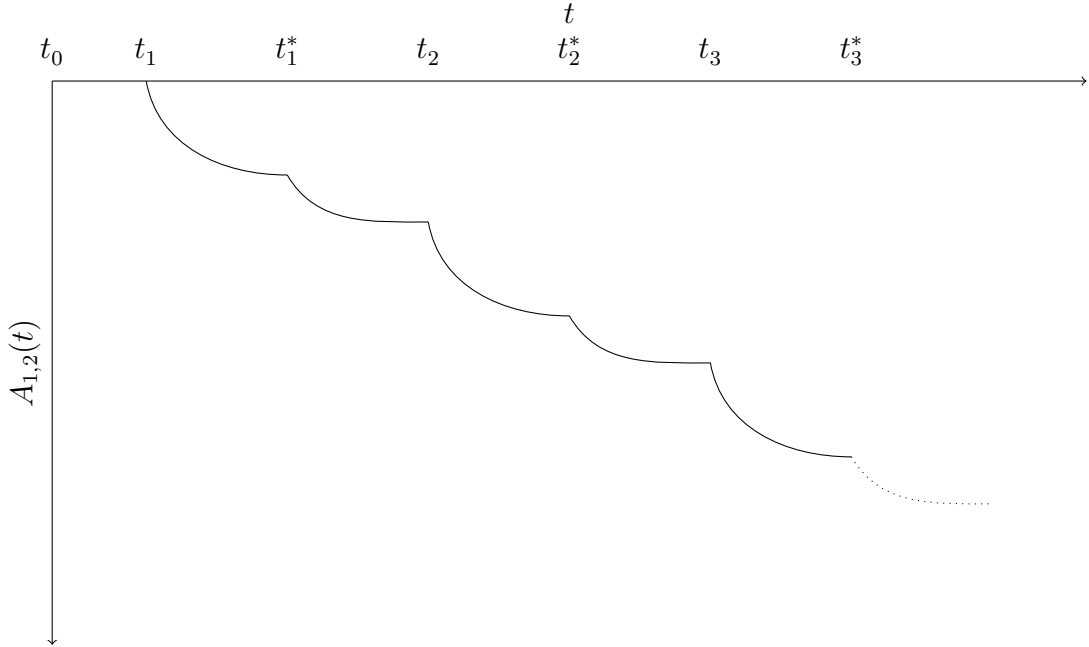

In Figure \ref{fig:5y} we illustrate possible forms of  $A_1(t)$ and  $A_2(t)$.   

\subsection{The total Cauchy stress in stress relaxation}
From the  results of this section we see that the total Cauchy stress $\bf T$ in a  material which   displays stress relaxation, softening and hysteresis is  given by
\begin{equation}
\mathbf{T}=\left\{\begin{array}{llll}
\phantom{\zeta_1(\lambda)\big\{}\textbf{T}^{\mathscr{E}}(\lambda),& \textrm{primary loading}, & t_0^{\phantom{*}}\leq t\leq t_1^{\phantom{*}}, & \textrm{path}\;\; P_0P_1\\[2mm]
\zeta_1(\lambda)\{\textbf{T}^{\mathscr{E}}(\lambda) + \textbf{T}^{\mathscr{R}}(\lambda, t)\},& \textrm{unloading}, & t_1^{\phantom{*}}\leq t\leq t_1^*, &  \textrm{path}\;\;P_1P_0\\[2mm]
\zeta_2(\lambda)\big\{\textbf{T}^{\mathscr{E}}(\lambda) + \textbf{T}^{\mathscr{R}}(\lambda, t)\big\},& \textrm{reloading}, & t^*_1\leq t\leq t_2^{\phantom{*}}, &  \textrm{path}\;\; P_0P_2\\[2mm]
\zeta_1(\lambda)\big\{\textbf{T}^{\mathscr{E}}(\lambda) + \textbf{T}^{\mathscr{R}}(\lambda, t)\big\},& \textrm{unloading}, & t_2^{\phantom{*}}\leq t\leq t_2^*, &  \textrm{path}\;\; P_2P_0\\[2mm]
\zeta_2(\lambda)\big\{\textbf{T}^{\mathscr{E}}(\lambda) + \textbf{T}^{\mathscr{R}}(\lambda, t)\big\},& \textrm{reloading}, & t^*_2\leq t\leq t_3^{\phantom{*}}, &  \textrm{path}\;\; P_0P_3\\[2mm]
 \;\; \dots&\;\; \dots&\;\; \dots&\;\; \dots
\end{array}\right. 
 \label{eq:16z}
 \end{equation}
 where $\textbf{T}^{\mathscr{E}}(\lambda)$ is the elastic stress (\ref{eq:5z}) and $ \textbf{T}^{\mathscr{R}}(\lambda, t)$ is the relaxation stress (\ref{eq:12z}).

The total stress (\ref{eq:16z}) falls to zero in $t>t_1$ and so we must have
$T^{\mathscr{R}}_{11}<0$ for $t>t_1$, implying that   $T^{\mathscr{R}}_{11}<0$ for $\lambda>1$.
Each of the quantities ${A}_0$, $  A_1(t)$, $  A_2(t)$ occurring in equation (\ref{eq:15z}) has positive coefficient for $\lambda>1$ and so at least one of them must be negative to maintain the requirement  $T^{\mathscr{R}}_{11}<0$ for $\lambda>1$.   In fact, we find in practice that all of them are negative.

Many authors have used the BKZ model and variations of it, though to our knowledge this is the first time it has been coupled with the Arruda-Boyce model to devize a model for stress-softening, hysteresis and stress relaxation.

\section{Residual strain}
\label{sec:residual}
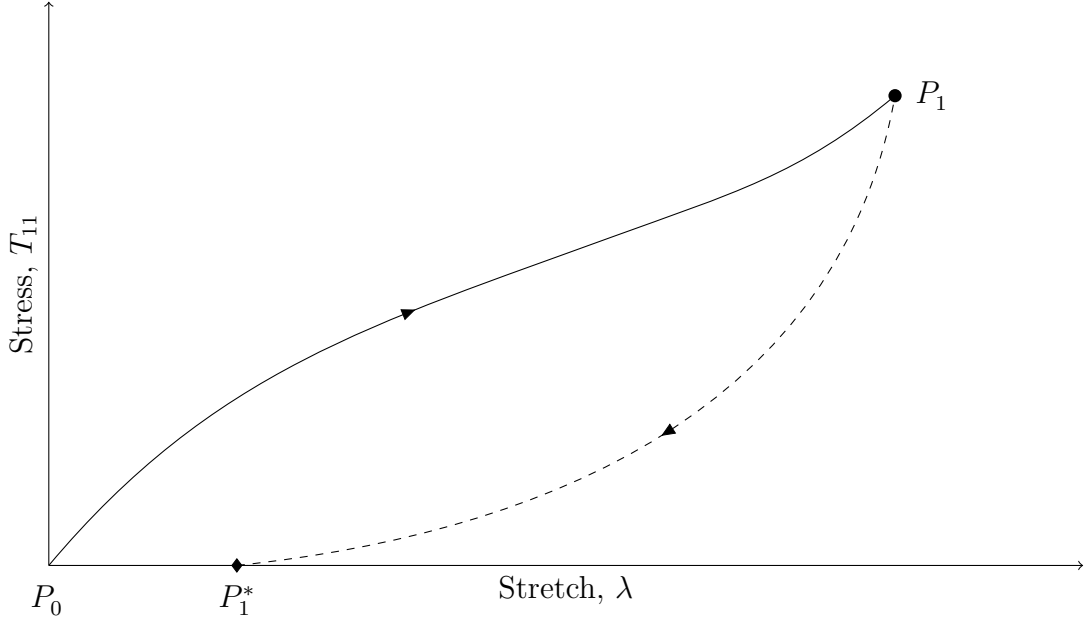
\begin{figure}[ht]
\centering
\begin{tikzpicture}[scale=1.25, decoration={
markings,
mark=at position 6cm with {\arrow[black]{triangle 45};},
mark=at position 19cm with {\arrowreversed[black]{triangle 45};},}
]
\draw[->] (0,0) -- (11,0)node[sloped,below,midway] {Stretch, $\lambda$} 
plot[only marks, mark=diamond*, mark size=2pt] coordinates {(2,0)};
\draw[->] (0,0) -- (0,6) node[sloped,above,midway] {Stress, $T_{11}$};
\fill (9,5) circle (2pt);
\draw [dashed]
	(2,0) to[out=6,in=260] node [sloped,below] {} (9,5);
\draw [postaction={decorate}][loosely dotted,line width=0.01pt]
	(0,0) to[out=50,in=200] node [sloped,above] {} (6,3.5)
(6,3.5) to[out=20,in=220] node [sloped,above] {} (9,5)
	(2,0) to[out=6,in=260] node [sloped,below] {} (9,5);
\draw [black]
	(0,0) to[out=50,in=200] node [sloped,above] {} (6,3.5)
(6,3.5) to[out=20,in=220] node [sloped,above] {} (9,5);
\coordinate [label=right:{$P_1^{\phantom{*}}$}] (B) at (9.1	,5);
\coordinate [label=below:{$P_0^{\phantom{*}}$}] (O) at (0,-0.1);
\coordinate [label=below:{$P^*_1$}] (O) at (2,-0.1);
\end{tikzpicture}
\caption{Residual strain in stress-softened rubber.}
\label{fig:6y}
\end{figure}

When a specimen of vulcanized rubber undergoes uniaxial tension it is observed that the unloading path might not return to the origin $P_0$ at zero stress but rather to a different point $P_1^*$, at strain $\lambda_1^*>1$, as indicated by the diamond marker in  Figure \ref{fig:6y}.  The unloading path $P_1^{\phantom{*}}P_1^*$ is shown as a dashed line
 in Figure \ref{fig:6y}.  
  The distance between the point $P_1^*$, where the unloading path reaches zero stress,  and the origin $P_0^{\phantom{*}}$, i.e. $\lambda_1^* - 1$, is  a measure of the increase in length of the material, that is, the degree of creep sustained by the material during unloading and reloading.

\subsection{The creep model of Bergstr\"om and Boyce}
\label{sec:BS}
\cite{bergstrom} introduced a continuum model of effective creep in order to explain the existence of residual strain.  Their model is based on the theoretical work of  \cite[page 213]{doi} which itself is rooted in statistical mechanics.  The length of a polymer chain is denoted by ${\lambda}_\chain$ and the \cite{arruda} eight-chain model is employed once more to deduce that for an isotropic material
\[
\lambda_\chain=\sqrt{\frac{I_1}{3}}.
\]
This equation has been used by \cite[eqn 22 and 23]{bergstrom} to derive 
the following equation which describes how the effective creep rate depends on the chain length, and so on the first invariant $I_1$ of  $\bf B$:
\begin{equation}
\mbox{effective creep rate} = c\left[{\lambda}_\chain - 1\right]^{-1}=c\left[\sqrt{\frac{I_1}{3}}-1\right]^{-1},
\label{eq:17z}
\end{equation}
where $c$ is a  constant. 

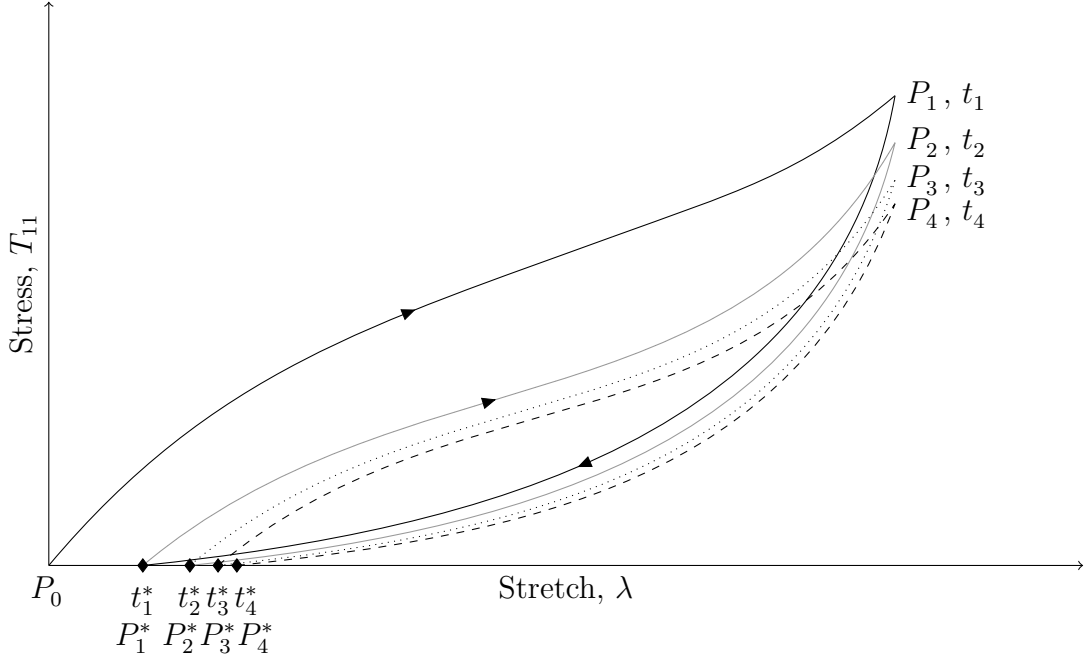
\begin{figure}[ht]
\centering
\begin{tikzpicture}[scale=1.25, decoration={
markings,
mark=at position 6cm with {\arrow[black]{triangle 45};},
mark=at position 19cm with {\arrowreversed[black]{triangle 45};},
mark=at position 31cm with {\arrow[black]{triangle 45};},}
]
\draw[->] (0,0) -- (11,0)node[sloped,below,midway] {Stretch, $\lambda$} ;
\draw[->] (0,0) -- (0,6) node[sloped,above,midway] {Stress, $T_{11}$};
\draw [black](0,0) to[out=50,in=200] node [sloped,above] {} (6,3.5);
\draw [black](6,3.5) to[out=20,in=220] node [sloped,above] {} (9,5);
\draw [black] (1.0,0) to [out=6,in=260] node [sloped,above] {} (9,5);
\draw [black!40] 
(1,0) to[out=40,in=240] node [sloped,above] {} (9,4.5)
(1.5,0) to[out=6,in=256.5] node [sloped,below] {} (9,4.5);
\draw [dotted]
(1.5,0) to[out=40,in=238] node [sloped,below] {} (9,4.1)
(1.8,0) to[out=6,in=253] node [sloped,above] {} (9,4.1);
\draw [dashed] 
(1.8,0) to[out=40,in=236] node [sloped,below] {} (9,3.85)
(2.0,0) to[out=6,in=250] node [sloped,above] {} (9,3.85);
\draw [postaction={decorate}][loosely dotted,line width=0.01pt]
(0,0) to[out=50,in=200] node [sloped,above] {} (6,3.5)
(6,3.5) to[out=20,in=220] node [sloped,above] {} (9,5)
	(1.0,0) to[out=6,in=260] node [sloped,above] {} (9,5)
	(1,0) to[out=40,in=240] node [sloped,above] {} (9,4.5)
	(1,0) to[out=40,in=240] node [sloped,above] {} (9,4.5);
\coordinate [label=right:{$P_1^{\phantom{*}}, \,t_1^{\phantom{*}}$}] (B) at (9.0	,5);
\coordinate [label=right:{$P_2^{\phantom{*}}, \,t_2^{\phantom{*}}$}] (B) at (9.0	,4.5);
\coordinate [label=right:{$P_3^{\phantom{*}}, \,t_3^{\phantom{*}}$}] (B) at (9.0	,4.1);
\coordinate [label=right:{$P_4^{\phantom{*}}, \,t_4^{\phantom{*}}$}] (B) at (9.0	,3.75);
\coordinate [label=below:{$t_1^*$}] (B) at (1,-0.1);
\coordinate [label=below:{$t_2^*$}] (B) at (1.5,-0.1);
\coordinate [label=below:{$t_3^*$}] (B) at (1.8,-0.1);
\coordinate [label=below:{$t_4^*$}] (B) at (2.1,-0.1);
\coordinate [label=below:{$P_1^*$}] (B) at (0.9,-0.5);
\coordinate [label=below:{$P_2^*$}] (B) at (1.4,-0.5);
\coordinate [label=below:{$P_3^*$}] (B) at (1.8,-0.5);
\coordinate [label=below:{$P_4^*$}] (B) at (2.2,-0.5);
\coordinate [label=below:{$ $}] (B) at (0.9,0);
\coordinate [label=below:{$ $}] (B) at (1.2,0);
\coordinate [label=below:{$ $}] (B) at (1.5,0);
\coordinate [label=below:{$P_0^{\phantom{*}}$}] (O) at (0,0);
\draw plot[only marks, mark=diamond*, mark size=2pt] coordinates {(1,0) (1.5,0) (1.8,0) (2,0)} ;
\end{tikzpicture}
\caption{Cyclic stress-softening.}
\label{fig:7y}
\end{figure}

The time dependence of a cyclically stretched rubber specimen  is illustrated  in Figure \ref{fig:7y}. Initially, the specimen is loaded to the  strain $\lambda=\lambda_\M$ at point $P_1^{\phantom{*}}$ and time $t_1^{\phantom{*}}$.  The material is then unloaded to zero stress at point $P_1^*$ and time $t_1^*$, at which point the strain is $\lambda_1^* >1$. Reloading  commences immediately at time $t_1^{*}$,  terminating at the  strain  $\lambda=\lambda_\M$ at point $P_2^{\phantom{*}}$ and time $t_2^{\phantom{*}}$.   The material is then unloaded to zero stress at point $P_2^*$ and time $t_2^*$, at which point the strain is $\lambda_2^* >\lambda_1^* >1$.  Reloading then  commences immediately at time $t_2^{*}$,  and so on. ¤  The points  $P^*_n$  are indicated by the diamond markers in Figure \ref{fig:7y}.  This pattern continues throughout the reloading and unloading process as the paths tend to equilibrium positions.  
Due to the inherent stress-softening features of the material the time intervals $t_1^* - t^{\phantom{*}}_1$ and  $t_2^* - t^{\phantom{*}}_2$  may  be unequal.  

\subsection{An extension of the model of Bergstr\"om and Boyce}
To predict the amount of residual strain accruing during cyclic unloading and reloading we replace the constant $c$ in equation (\ref{eq:17z}) by a time dependent function $c(t)$ such as 
\begin{equation}
c(t)= d\left[1 + \left[\tanhs\breve{a}(t)\right]^{a_1}\right],
\label{eq:18z}
\end{equation}
where $a_1>0$ and  $d$ are  material constants and $\breve{a}(t)>0$ is a continuous, increasing function of time. 
As $t \to \infty$ we see that $c(t) \to 2d$ so that the unloading and reloading paths tend to equilibrium values.
This function is capable of representing experimental data for suitable choice of $d$, $a_1$ and $\breve{a}(t)$ . 
The constant $d$ is selected to ensure that the first  unloading path $P_1^{\phantom{*}}P_1^*$ ceases at the point $P_1^*$. 

We assume that creep leading to residual strain does not occur during primary loading but evolves throughout the unloading and reloading process, though this creep  may run at different rates on the  unloading and  reloading paths.

The function $a(t)$ is defined  by
\begin{equation}
a(t)=\left\{\begin{array}{llll}
   0                                      & \textrm{primary loading}, & t_0^{\phantom{*}}\leq t\leq t_1^{\phantom{*}}, & \textrm{path}\;\; P_0^{\phantom{*}}P_1^{\phantom{*}}\\[0.5mm]
  \breve{a}(\Phi_1(t-t_1))& \textrm{unloading}, & t_1^{\phantom{*}}\leq t\leq t_1^*, &  \textrm{path}\;\; P_1^{\phantom{*}}P_1^*\\[0.5mm]
  \breve{a}(\Phi_2(t-t_1))& \textrm{reloading}, & t^*_1\leq t\leq t_2^{\phantom{*}}, &  \textrm{path}\;\; P_1^*P_2^{\phantom{*}}\\[0.5mm]
  \breve{a}(\Phi_1(t-t_1))& \textrm{unloading}, & t_2^{\phantom{*}}\leq t\leq t_2^*, &  \textrm{path}\;\; P_2^{\phantom{*}}P_2^*\\[0.5mm]
  \breve{a}(\Phi_2(t-t_1))& \textrm{reloading}, & t^*_2\leq t\leq t_3^{\phantom{*}}, &  \textrm{path}\;\; P_2^*P_3^{\phantom{*}}\\[0.5mm]
 \;\; \dots&\;\; \dots&\;\; \dots&\;\; \dots
\end{array}\right.
\label{eq:19z}
\end{equation}
with $\Phi_1$ and $\Phi_2$ being continuous functions of time. The form of $a(t)$ is similar to that of  $A_{1, 2}(t)$ shown in Figure \ref{fig:5y}.

In the current model it is assumed that the specimen of rubber is being cyclically stretched to the same final strain 
$\lambda_\M$. During each  cycle the residual stretch $\lambda^*_n$ of the unloaded specimen at point $P^*_n$  increases until an equilibrium state is reached. By combining equations (\ref{eq:17z}) and (\ref{eq:18z}) we obtain an expression for the Cauchy stress when the material is undergoing a number of unloading and reloading cycles:
\begin{equation}
\textbf{T}^{\mathscr{C}}(\lambda, t)=-p \textbf{I}+\left\{d\left[\sqrt{\frac{I_1}{3}}-1\right]^{-1}\left[1+\left[\tanhs a(t)\right]^{a_1}\right]\right\} \textbf{B}\quad\mbox{for}\;\; t>t_1 \;\; \mbox{and}\;\; \lambda>1.
\label{eq:20z}
\end{equation}
The number of cycles is accounted for in equation (\ref{eq:20z})  by the definition (\ref{eq:19z}) of $a(t)$.

For $t<t_1$ and  $\lambda=1$, $\textbf{T}^{\mathscr{C}}(\lambda, t)$ must vanish as there is no residual strain. 
The singularity of equation   (\ref{eq:20z}) at  $\lambda=1$ is not relevant as the residual strains all satisfy $\lambda^*_n>1$.

From equation (\ref{eq:20z}) applied to uniaxial tension, after eliminating $p$, we obtain
\begin{equation}
{T}^{\mathscr{C}}_{11}(\lambda, t)=d\left[\lambda^2-\frac{1}{\lambda}\right]\left[\sqrt{\frac{I_1}{3}}-1\right]^{-1}\left[1+\left[\tanhs a(t)\right]^{a_1}\right].
\label{eq:21z}
\end{equation}
The total stress (\ref{eq:21z}) falls to zero in $t>t^*_1$ and so we must have
$T^{\mathscr{C}}_{11}<0$ for $t>t^*_1$, implying that   $T^{\mathscr{C}}_{11}<0$ for $\lambda>1$.
This requirement forces $d<0$ in equation (\ref{eq:21z}).
 
It follows that  the total Cauchy stress $\bf T$ in a  material displaying stress relaxation,  softening, hysteresis and residual strain is  given by
\begin{equation}
\mathbf{T}=\left\{\begin{array}{llll}
\phantom{\zeta_2(\lambda)\,}\textbf{T}^{\mathscr{E}}(\lambda),& \textrm{primary loading}, & t_0\leq t\leq t_1, & \textrm{path}\;\; P_0^{\phantom{*}}P_1^{\phantom{*}}  \\[2mm]
\zeta_1\,(\lambda)\textbf{T}^{\mathscr{E} + \mathscr{R} + \mathscr{C} }(\lambda, t)
,& \textrm{unloading}, & t_1^{\phantom{*}}\leq t\leq t_1^{*}, &
\textrm{path}\;\; P_1^{\phantom{*}}P_1^{{*}} \\[2mm]
\zeta_2\,(\lambda)\textbf{T}^{\mathscr{E} + \mathscr{R} + \mathscr{C} }(\lambda, t),& \textrm{reloading}, & t^{*}_1\leq t\leq t_2^{\phantom{*}}, 
&  \textrm{path}\;\; P_1^{{*}}P_2^{\phantom{*}}  \\[2mm]
\zeta_1\,(\lambda)\textbf{T}^{\mathscr{E} + \mathscr{R} + \mathscr{C} }(\lambda, t),& \textrm{unloading}, & t_2^{\phantom{*}}\leq t\leq t_2^{*}, 
&   \textrm{path}\;\; P_2^{\phantom{*}}P_2^{{*}}  \\[2mm]
\zeta_2\,(\lambda)\textbf{T}^{\mathscr{E} + \mathscr{R} + \mathscr{C} }(\lambda, t),& \textrm{reloading}, & t^{*}_2\leq t\leq t_3^{\phantom{*}}, 
&   \textrm{path}\;\; P_2^{{*}}P_3^{\phantom{*}}  \\[2mm]
 \;\; \dots&\;\; \dots&\;\; \dots&\;\;\dots
\end{array}\right. 
\label{eq:22z}
\end{equation}
in which for notational convenience we have defined a  stress
\begin{equation}
  \textbf{T}^{\mathscr{E} + \mathscr{R} + \mathscr{C} }(\lambda, t)  = \textbf{T}^{\mathscr{E}}(\lambda) + \textbf{T}^{\mathscr{R}}(\lambda, t) + \textbf{T}^{\mathscr{C}}(\lambda, t),
\label{eq:23z}
\end{equation}
where $\textbf{T}^{\mathscr{E}}(\lambda)$, $\textbf{T}^{\mathscr{R}}(\lambda, t)$ and $\textbf{T}^{\mathscr{C}}(\lambda, t)$ are given by equations   (\ref{eq:5z}),  (\ref{eq:12z}) and  (\ref{eq:20z}), respectively.

\section{Creep of residual strain}
\label{sec:creepresidual}

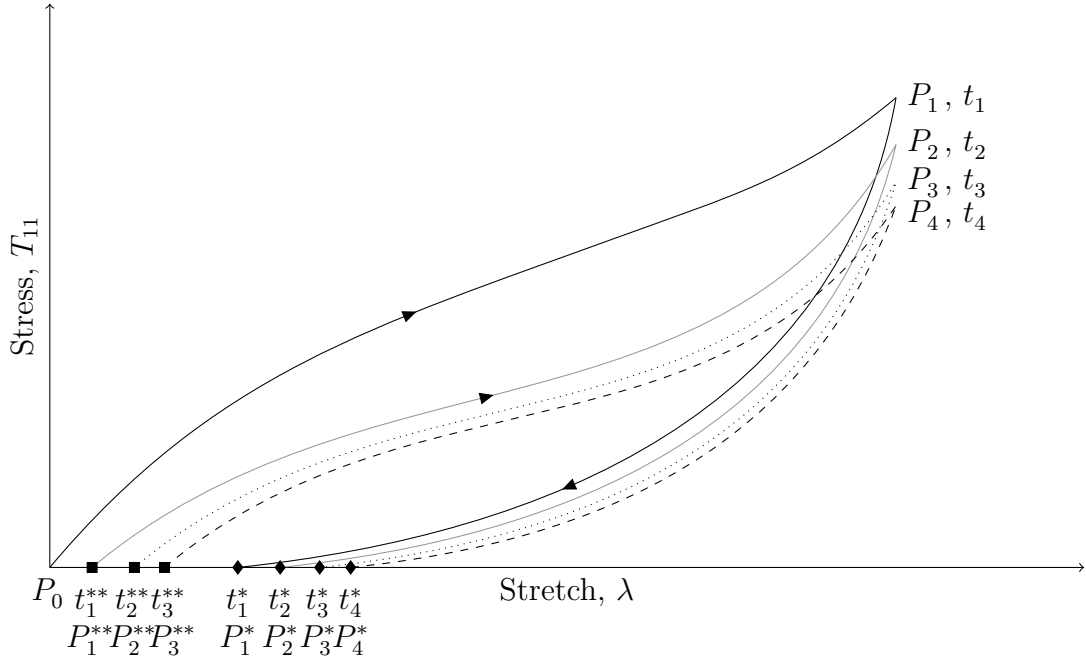
\begin{figure}[ht]
\centering
\begin{tikzpicture}[scale=1.25, decoration={
markings,
mark=at position 6cm with {\arrow[black]{triangle 45};},
mark=at position 17.5cm with {\arrowreversed[black]{triangle 45};},
mark=at position 30.5cm with {\arrow[black]{triangle 45};},}
]
\draw[->] (0,0) -- (11,0)node[sloped,below,midway] {Stretch, $\lambda$} ;
\draw[->] (0,0) -- (0,6) node[sloped,above,midway] {Stress, $T_{11}$};
\draw [black](0,0) to[out=50,in=200] node [sloped,above] {} (6,3.5);
\draw [black](6,3.5) to[out=20,in=220] node [sloped,above] {} (9,5);
\draw [black] (2,0) to [out=6,in=260] node [sloped,above] {} (9,5);
\draw [black!40] 
(0.45,0) to[out=40,in=240] node [sloped,above] {} (9,4.5)
(2.45,0) to[out=6,in=256.5] node [sloped,below] {} (9,4.5);
\draw [dotted]
(0.9,0) to[out=40,in=238] node [sloped,below] {} (9,4.1)
(2.87,0) to[out=6,in=253] node [sloped,above] {} (9,4.1);
\draw [dashed] 
(1.22,0) to[out=40,in=236] node [sloped,below] {} (9,3.85)
(3.2,0) to[out=6,in=250] node [sloped,above] {} (9,3.85);
\draw [postaction={decorate}][loosely dotted,line width=0.01pt]
(0,0) to[out=50,in=200] node [sloped,above] {} (6,3.5)
(6,3.5) to[out=20,in=220] node [sloped,above] {} (9,5)
	(2,0) to[out=6,in=260] node [sloped,above] {} (9,5)
	(0.45,0) to[out=40,in=240] node [sloped,above] {} (9,4.5)
	(0.45,0) to[out=40,in=240] node [sloped,above] {} (9,4.5);
\coordinate [label=right:{$P_1^{\phantom{*}}, \,t_1^{\phantom{*}}$}] (B) at (9.0	,5);
\coordinate [label=right:{$P_2^{\phantom{*}}, \,t_2^{\phantom{*}}$}] (B) at (9.0	,4.5);
\coordinate [label=right:{$P_3^{\phantom{*}}, \,t_3^{\phantom{*}}$}] (B) at (9.0	,4.1);
\coordinate [label=right:{$P_4^{\phantom{*}}, \,t_4^{\phantom{*}}$}] (B) at (9.0	,3.75);
\coordinate [label=below:{$t_1^*$}] (B) at (2,-0.1);
\coordinate [label=below:{$t_2^*$}] (B) at (2.45,-0.1);
\coordinate [label=below:{$t_3^*$}] (B) at (2.85,-0.1);
\coordinate [label=below:{$t_4^*$}] (B) at (3.2,-0.1);
\coordinate [label=below:{$t_1^{**}$}] (B) at (0.45,-0.1);
\coordinate [label=below:{$t_2^{**}$}] (B) at (0.88,-0.1);
\coordinate [label=below:{$t_3^{**}$}] (B) at (1.26,-0.1);
\coordinate [label=below:{$P_1^*$}] (B) at (2,-0.5);
\coordinate [label=below:{$P_2^*$}] (B) at (2.45,-0.5);
\coordinate [label=below:{$P_3^*$}] (B) at (2.85,-0.5);
\coordinate [label=below:{$P_4^*$}] (B) at (3.2,-0.5);

\coordinate [label=below:{$P_1^{**}$}] (B) at (0.43,-0.5);
\coordinate [label=below:{$P_2^{**}$}] (B) at (0.88,-0.5);
\coordinate [label=below:{$P_3^{**}$}] (B) at (1.3,-0.5);

\coordinate [label=below:{$ $}] (B) at (0.9,0);
\coordinate [label=below:{$ $}] (B) at (1.2,0);
\coordinate [label=below:{$ $}] (B) at (1.5,0);
\coordinate [label=below:{$P_0^{\phantom{*}}$}] (O) at (0,0);
\draw plot[only marks, mark=diamond*, mark size=2pt] coordinates {(2,0) (2.45,0) (2.87,0) (3.2,0)} ;
\draw plot[only marks, mark=square*, mark size=1.5pt] coordinates {(0.45,0) (0.9,0) (1.22,0) } ;
\end{tikzpicture}
\caption{Cyclic stress-softening with residual strain.}
\label{fig:8y}
\end{figure}

Suppose now that reloading does not commence at the same time $t^*_n$ that unloading ceased, as was the case in Figure \ref{fig:7y}.  Instead, the material that was fully unloaded at time $t^*_n$ and stretch $\lambda_n^{*}$ is now left stress free until the later time $t^{**}_n$ when the point $P^{**}_n$ is reached, with stretch $\lambda_n^{**}$, satisfying $1<\lambda_n^{**}<\lambda_n^{*}$.   Then at time $t^{**}_n$ reloading  recommences and proceeds until the point  $P_{n+1}$ is reached.
This new phenomenon is illustrated in  Figure \ref{fig:8y} where diamond markers indicate where unloading ceases (as before) and square markers indicate the new points where reloading commences.

It is observed experimentally that during cyclic unloading and reloading the relaxation paths progressively move away from the first unloading and reloading paths until an equilibrium state is reached, as illustrated in Figure \ref{fig:8y}, and highlighted there by the  diamond and square markers  moving to the right as time progresses.    Due to the inherent stress-softening features of the material the time intervals 
$t_1^* - t^{\phantom{*}}_1$ and $t_2^{*} - t^{\phantom{*}}_2$ may be unequal as also may  $t_1^{**} - t^{\phantom{*}}_1$ and $t_2^{**} - t^{\phantom{*}}_2$.

The stress-relaxation functions $A_{1, 2}(t)$ and residual strain function $a(t)$  operate during  unloading and reloading and also for the time periods $t^{*}_n\leq t \leq t^{**}_n$ of zero stress.
During these stress-free time periods  the material continues to undergo stress relaxation.   Thus equation (\ref{eq:14z})  is replaced by
\begin{equation}
A_{1, 2}(t)=\left\{\begin{array}{llll}
   0                                      & \textrm{primary loading}, & t_0^{\phantom{*}}\leq t\leq t_1^{\phantom{*}}, &  \textrm{path}\;\; P_0^{\phantom{*}}P_1^{\phantom{*}}\\[2mm]
\breve{A}_{1, 2}(\phi_1(t-t_1))& \textrm{unloading}, & t_1^{\phantom{*}}\leq t\leq t_1^*, &  \textrm{path}\;\; P_1^{\phantom{*}}P_1^*\\[2mm]
\breve{A}_{1, 2}(\phi_1(t-t_1))& \textrm{stress free}, & t^*_1\leq t\leq t_1^{**}, &  \textrm{path}\;\; P_1^*P_1^{**}\\[2mm] 
\breve{A}_{1, 2}(\phi_2(t-t_1))& \textrm{reloading}, & t^{**}_1\leq t\leq t_2^{\phantom{*}}, &  \textrm{path}\;\; P_1^{**}P_2^{\phantom{*}}\\[2mm]
\breve{A}_{1, 2}(\phi_1(t-t_1))& \textrm{unloading}, & t_2^{\phantom{*}}\leq t\leq t_2^*, &  \textrm{path}\;\; P_2^{\phantom{*}}P_2^*\\[2mm]
\breve{A}_{1, 2}(\phi_1(t-t_1))& \textrm{stress free}, & t^{*}_2\leq t\leq t^{**}_2, &  \textrm{path}\;\; P_2^*P_2^{**}\\[2mm]
\breve{A}_{1, 2}(\phi_2(t-t_1))& \textrm{reloading}, & t^{**}_1\leq t\leq t_3^{\phantom{*}}, &  \textrm{path}\;\; P_2^{**}P_3^{\phantom{*}}\\[2mm]
 \;\; \dots&\;\; \dots&\;\; \dots&\;\; \dots 
\end{array}\right.
\label{eq:24z}
\end{equation}
where $\phi_1$ and $\phi_2$ are as in (\ref{eq:14z}).
For simplicity,  on the stress-free paths we  employ $\phi_1$ as the argument for  $A_{1, 2}(t)$ as we did on the unloading paths. 

The residual strain function $a(t)$, defined by equation (\ref{eq:19z}), is replaced by
\begin{equation}
a(t)=\left\{\begin{array}{llll}
   0                                      & \textrm{primary loading}, & t_0^{\phantom{*}}\leq t\leq t_1^{\phantom{*}}, &  \textrm{path}\;\; P_0^{\phantom{*}}P_1^{\phantom{*}}\\[2mm]
\breve{a}(\Phi_1(t-t_1))& \textrm{unloading}, & t_1^{\phantom{*}}\leq t\leq t_1^*, &  \textrm{path}\;\; P_1^{\phantom{*}}P_1^*\\[2mm]
\breve{a}(\Phi_1(t-t_1))& \textrm{stress free}, & t^*_1\leq t\leq t_1^{**}, &  \textrm{path}\;\; P_1^*P_1^{**}\\[2mm] 
\breve{a}(\Phi_2(t-t_1))& \textrm{reloading}, & t^{**}_1\leq t\leq t_2^{\phantom{*}}, &  \textrm{path}\;\; P_1^{**}P_2^{\phantom{*}}\\[2mm]
\breve{a}(\Phi_1(t-t_1))& \textrm{unloading}, & t_2^{\phantom{*}}\leq t\leq t_2^*, &  \textrm{path}\;\; P_2^{\phantom{*}}P_2^*\\[2mm]
\breve{a}(\Phi_1(t-t_1))& \textrm{stress free}, & t^{*}_2\leq t\leq t^{**}_2, &  \textrm{path}\;\; P_2^*P_2^{**}\\[2mm]
\breve{a}(\Phi_2(t-t_1))& \textrm{reloading}, & t^{**}_1\leq t\leq t_3^{\phantom{*}}, &  \textrm{path}\;\; P_2^{**}P_3^{\phantom{*}}\\[2mm]
 \;\; \dots&\;\; \dots&\;\; \dots&\;\; \dots
\end{array}\right.
\label{eq:25z}
\end{equation}
where $\Phi_1$ and $\Phi_2$ are as in equation  (\ref{eq:19z}).
For simplicity,  on the stress-free paths we  employ $\Phi_1$ as the argument for  $a(t)$ as we did on the unloading paths. 

In order to account for any change in residual strain during the time periods $t^{*}_n\leq t \leq t^{**}_n$,  the 
stress-free range in equation (\ref{eq:25z}),  we replace the constant $d$ in equation (\ref{eq:21z})  by $d_\omega$,  allowing it to take a value  $d_1$ for  unloading and a different value $d_2$ for  reloading.  Then equation (\ref{eq:20z})  becomes
\begin{equation}
\textbf{T}^{\mathscr{C}}(\lambda, t)=-p \textbf{I}+\left\{d_\omega\left[\sqrt{\frac{I_1}{3}}-1\right]^{-1}\left[1+\left[\tanhs a(t)\right]^{a_1}\right]\right\} \textbf{B}\quad\mbox{for}\;\;  t > t_1 \;\; \mbox{and}\;\; \lambda>1 ,
\label{eq:26z}
\end{equation}
and the corresponding uniaxial stress ${T}^{\mathscr{C}}_{11}(\lambda, t)$ is  given by equation (\ref{eq:21z}) with $d$ replaced by $d_\omega$.

\section{Constitutive model}
\label{sec:constitutive}

From the results of the previous section it follows that  the total Cauchy stress $\bf T$ in a  material displaying  softening, hysteresis, stress relaxation,  residual strain and creep of residual strain is  given by
\begin{equation}
\mathbf{T}=\left\{\begin{array}{llll}
 \phantom{\zeta_1(\lambda)\,}\textbf{T}^{\mathscr{E}}(\lambda),& \textrm{primary loading}, & t_0^{\phantom{*}}\leq t\leq t_1^{\phantom{*}}, & \textrm{path}\;\; P_0^{\phantom{*}}P_1^{\phantom{*}}\\[2mm]                 
\zeta_1(\lambda)\, \textbf{T}^{\mathscr{E} + \mathscr{R} + \mathscr{C} }(\lambda, t) ,& \textrm{unloading}, & t_1^{\phantom{*}}\leq t\leq t_1^{*}, &  \textrm{path}\;\; P_1^{\phantom{*}}P_1^*\\[2mm]
\phantom{\zeta_1(\lambda)\,}{\bf 0}, & \textrm{stress free}, & t^*_1\leq t\leq t_1^{**}, &  \textrm{path}\;\; P_1^*P_1^{**}\\[2mm] 
\zeta_2(\lambda)\, \textbf{T}^{\mathscr{E} + \mathscr{R} + \mathscr{C} }(\lambda, t) ,& \textrm{reloading}, & t^{**}_1\leq t\leq t_2^{\phantom{*}}, &  \textrm{path}\;\; P_1^{**}P_2^{\phantom{*}}\\[2mm]
\zeta_1(\lambda)\, \textbf{T}^{\mathscr{E} + \mathscr{R} + \mathscr{C} }(\lambda, t) ,& \textrm{unloading}, & t_2^{\phantom{*}}\leq t\leq t_2^{*}, &  \textrm{path}\;\; P_2^{\phantom{*}}P_2^*\\[2mm]
\phantom{\zeta_1(\lambda)\,}{\bf 0}, & \textrm{stress free}, & t^*_2\leq t\leq t_2^{**}, &  \textrm{path}\;\; P_2^*P_2^{**}\\[2mm] 
\zeta_2(\lambda)\, \textbf{T}^{\mathscr{E} + \mathscr{R} + \mathscr{C} }(\lambda, t) ,& \textrm{reloading}, & t^{**}_2\leq t\leq t_3^{\phantom{*}}, &  \textrm{path}\;\; P_2^{**}P_3^{\phantom{*}}\\[2mm]
 \;\; \dots&\;\; \dots&\;\; \dots&\;\; \dots
\end{array}\right. 
 \label{eq:27z}
 \end{equation}
 where once again the  stress  $ \textbf{T}^{\mathscr{E} + \mathscr{R} + \mathscr{C} }(\lambda, t) $, defined by (\ref{eq:23z}), is employed for notational convenience, except that here 
 $\textbf{T}^{\mathscr{C}}(\lambda, t)$ is given by  equation (\ref{eq:26z}) rather than by equation (\ref{eq:20z}).

To the best of our knowledge the effects of residual strain in relation to the Mullins effect during cyclic loading have not previously been considered in the literature  and so the resulting equation (\ref{eq:27z}) for the stress has not previously been exhibited.

On substituting the individual stresses  given by equations  (\ref{eq:6z}), (\ref{eq:15z}) and (\ref{eq:26z}) into equation (\ref{eq:27z}),   we obtain the following constitutive model for the Cauchy stress $\bf T$ in an incompressible isotropic  solid material that  models stress-softening, hysteresis, stress relaxation,    residual strain and creep of residual strain:
\begin{align}
\textbf{T} = & \, \left[1-\frac{1}{r_\omega}\left\{\tanhs\left(\frac{W_\M-W}{\mu b_\omega}\right)\right\}^{1/{\vartheta_\omega}}\right] \times\nonumber\\
& \times \Bigg\{ -p\textbf{I}+\mu\sqrt{\frac{N}{3I_1}}\mathscr{L}^{-1}\left(\sqrt{\frac{I_1}{3{{N}}}}\right)\textbf{B}\nonumber\\
&\qquad+\left[{A}_0\textbf{B}+\frac{1}{2}{A_1}(t)(I_1-3)\textbf{B} + {A_2}(t)(\textbf{B}^2-\textbf{B})\right]\nonumber\\
&\qquad+d_\omega\left[\sqrt{\frac{I_1}{3}}-1\right]^{-1}\left[1+\left[\tanhs a(t)\right]^{a_1}\right] \textbf{B}\Bigg\}.
\label{eq:28z}
\end{align}
As before, we eliminate $p$ from equation (\ref{eq:28z}) to obtain the uniaxial tension
\begin{align}
T_{11}= &\,\left[1-\frac{1}{r_\omega}\left\{\tanhs\left(\frac{W_\M-W}{\mu b_\omega}\right)\right\}^{1/{\vartheta_\omega}}\right] \times\nonumber \\
& \times \Bigg\{\mu (\lambda^2-\lambda^{-1})  \sqrt{\frac{N}{3I_1}}\mathscr{L}^{-1}\left(\sqrt{\frac{I_1}{3{{N}}}}\right)\nonumber\\
&\qquad+ (\lambda^2-\lambda^{-1}) \left[{A}_0+\frac{1}{2}{{A}_1(t)}(\lambda - 1)^2(1+2\lambda^{-1})+{{A}_2(t)}(\lambda^2-1+\lambda^{-1})\right]\nonumber \\
&\qquad+d_\omega (\lambda^2-\lambda^{-1}) \left[\sqrt{\frac{I_1}{3}}-1\right]^{-1}\left[1+\left[\tanhs a(t)\right]^{a_1}\right]\Bigg\}.
\label{eq:29z}
\end{align}

\tcr{The authors believe that this constitutive equation for cyclic stress-softening in  the Mullins effect is the first to incorporate all the inelastic effects of hysteresis, stress-relaxation, residual strain and creep of residual strain.}

\section{Numerical examples and comparison with experimental data}
\label{sec:numerical}
\begin{figure}[ht]
\centering
\begin{tikzpicture}
\node (0,0) {\includegraphics{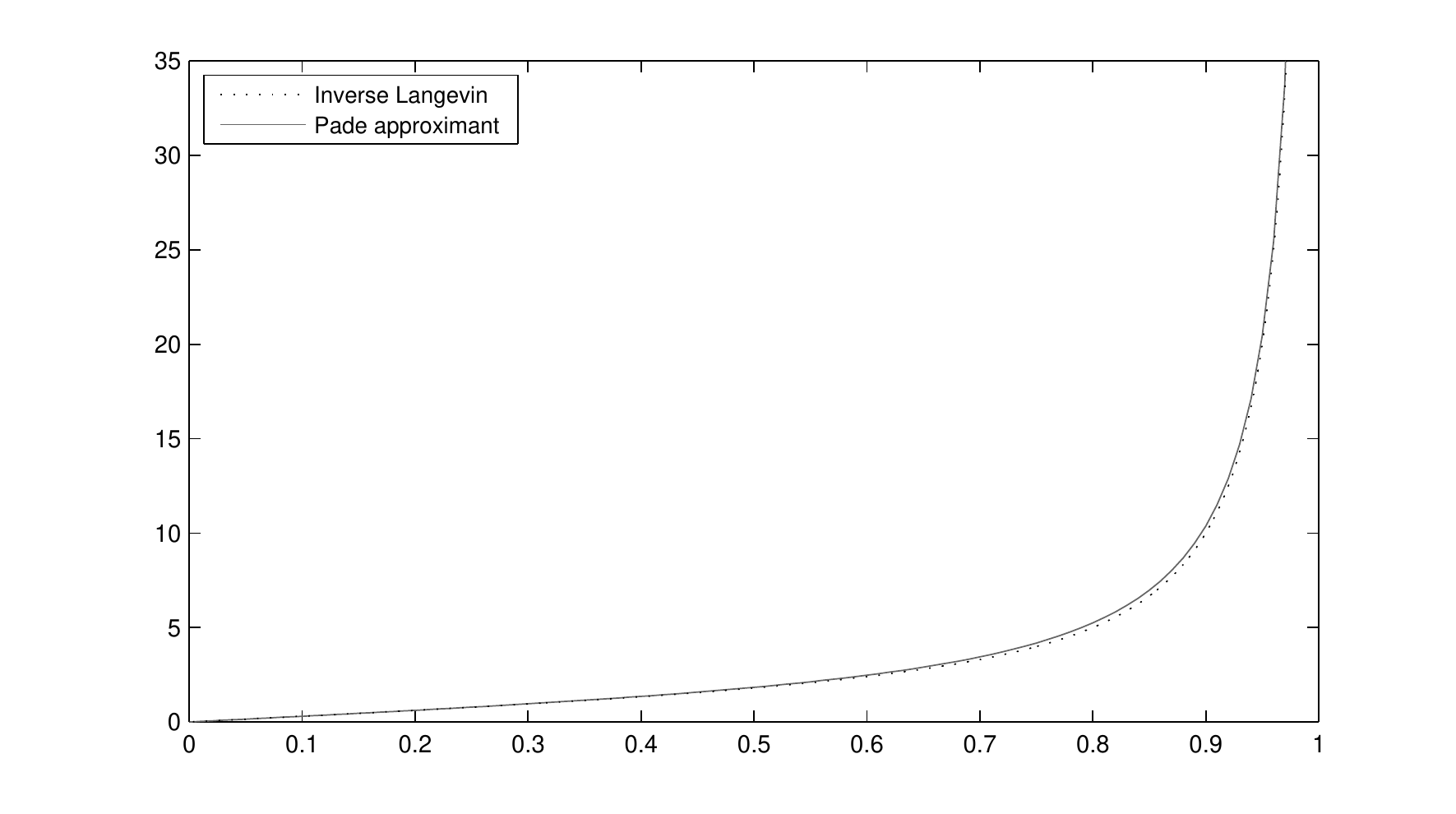}};
\draw  (-7.5,0.1) node [rotate=90] {\fontsize{10}{10} $\mathscr{L}^{-1}(x)$};
\draw  (-4.61,3.47) node {\fontsize{10}{10} $\acute{\phantom{e}}$};
\draw  (0.3,-4.6) node {\fontsize{10}{10} $x$};
\end{tikzpicture} 
\vspace{-20pt}
\caption{Comparison between the inverse Langevin function and the Pad\'e approximant.}
\label{fig:9y}
\end{figure}
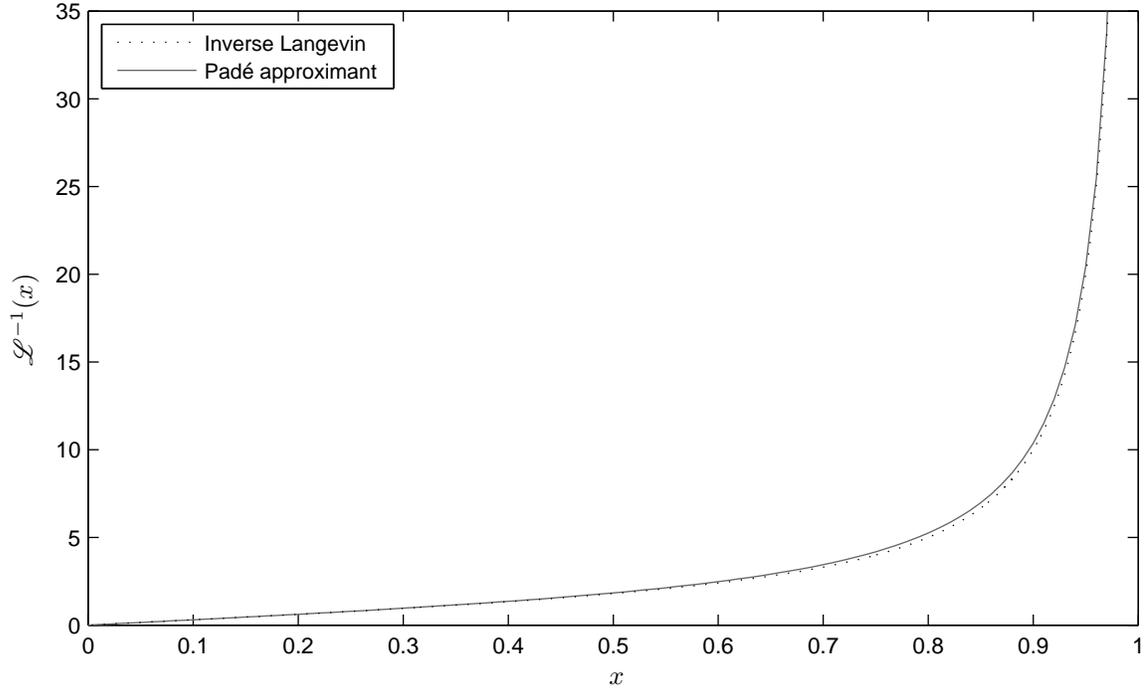

In our numerical work we  approximate the inverse Langevin function by its Pad\'e approximant  derived by \cite{cohen}, namely,
\begin{equation}
\mathscr{L}^{-1}(x)\approx\frac{3x-x^3}{1-x^2}.
\label{eq:30z}
\end{equation}
Figure \ref{fig:9y} illustrates the close agreement between the inverse Langevin function and its approximation (\ref{eq:30z}).

In our modelling of the Mullins effect we have used the Biot stress $T_{B11}$ defined by
\[
T_{B11}=\lambda^{-1}T_{11}
\]
in order to facilitate comparison with  experimental work.

\subsection{Cyclic stress-softening for certain typical materials}

\begin{figure}[ht]
\centering
\includegraphics[width=18cm,height=10cm]{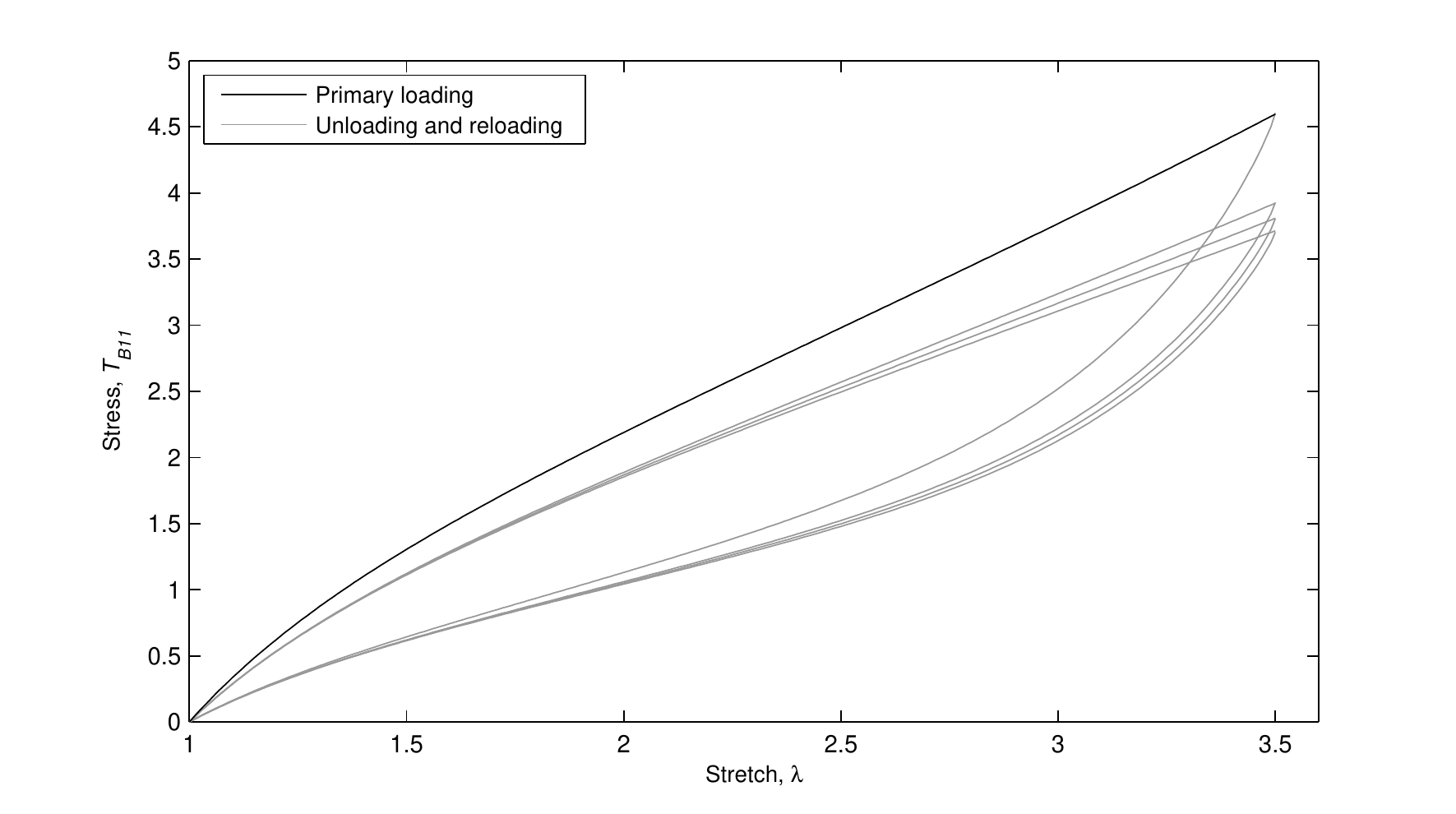}
\vspace{-20pt}
\caption{Cyclic stress-softening without residual strain.}
\label{fig:10y}
\end{figure}

\begin{figure}[ht]
\centering
\includegraphics[width=18cm,height=10cm]{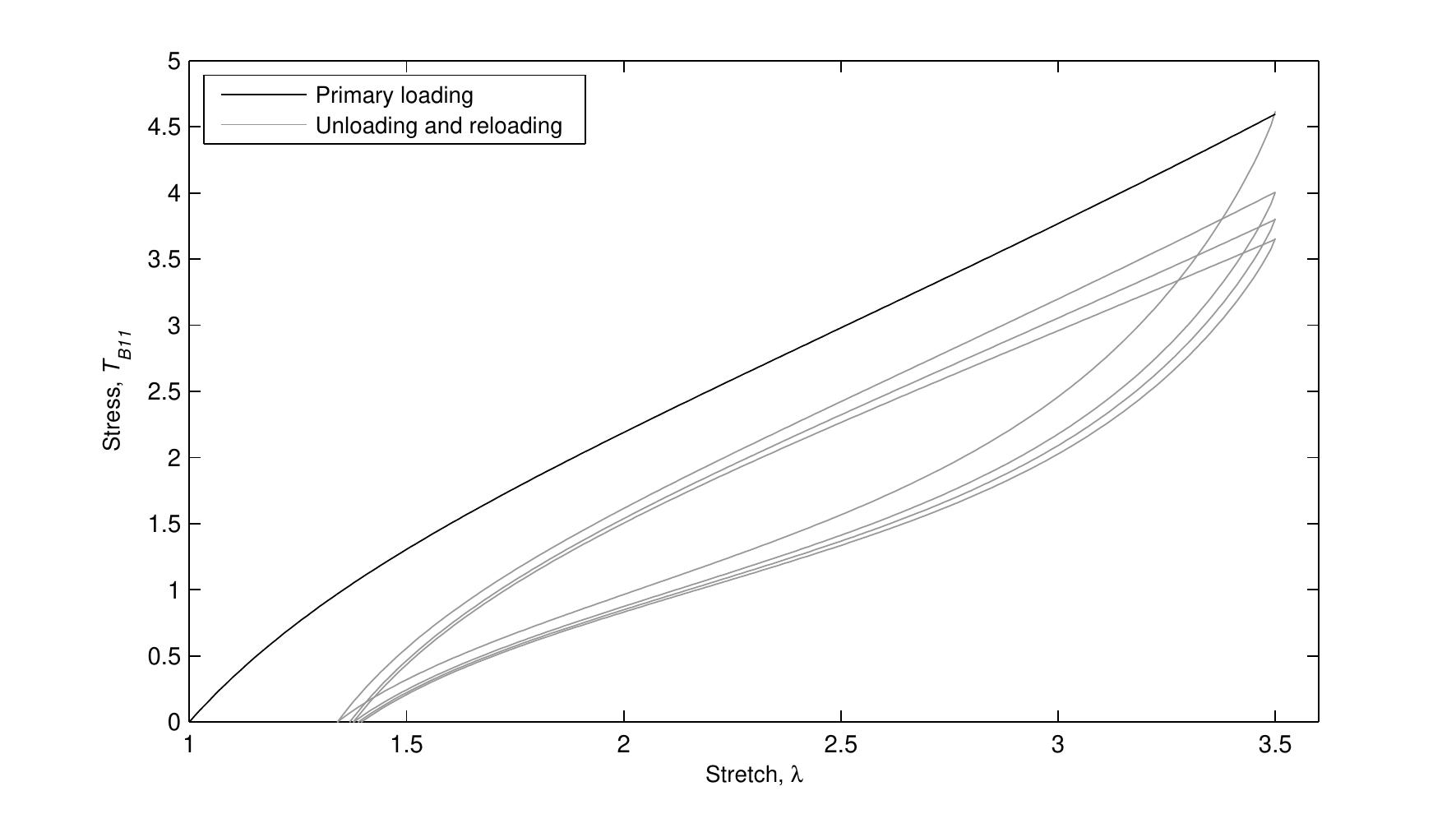}
\vspace{-20pt}
\caption{Cyclic stress-softening with residual strain.}
\label{fig:11y}
\end{figure}

\begin{figure}[ht]
\centering
\includegraphics[width=18cm,height=10cm]{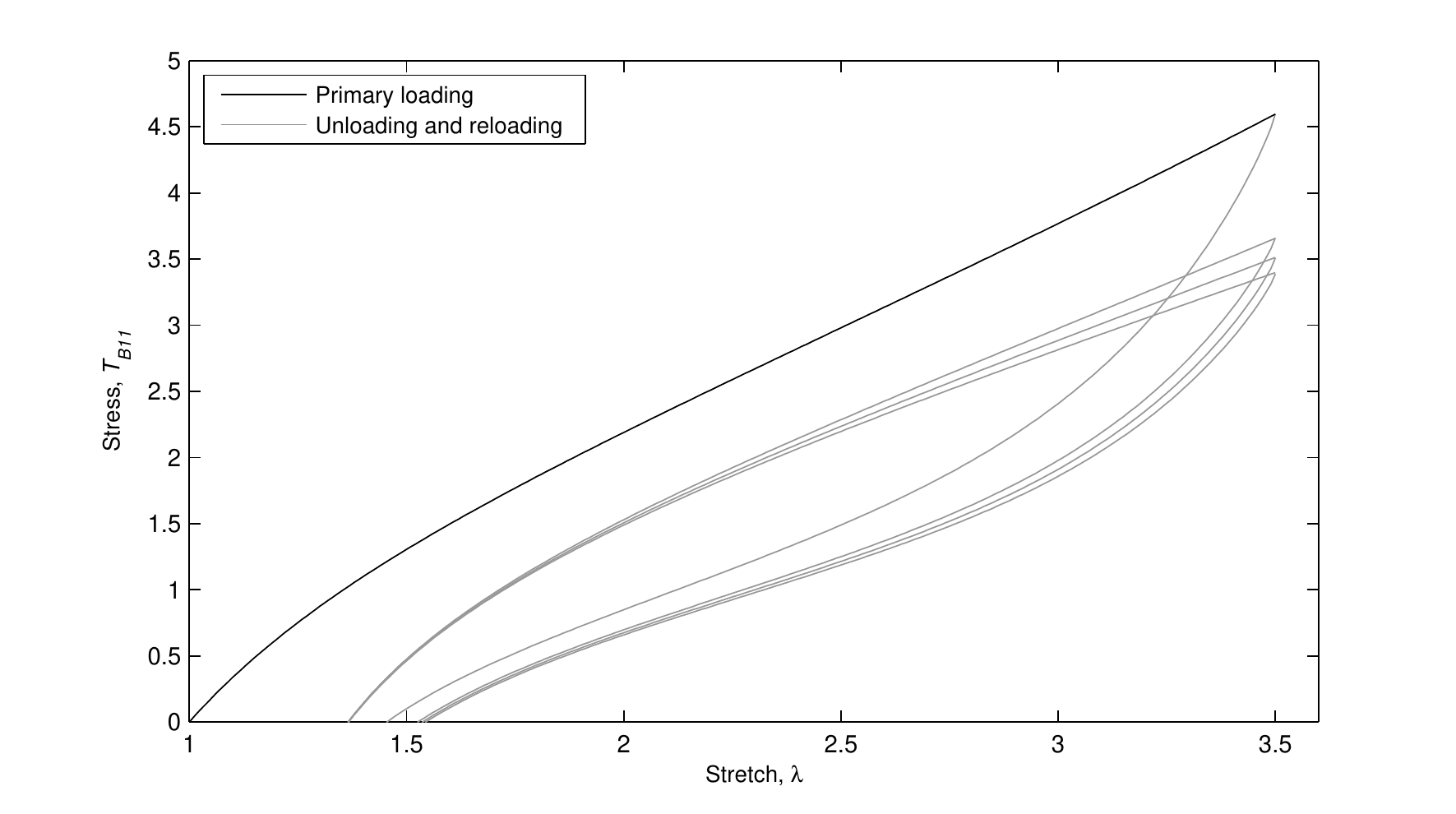}
\vspace{-20pt}
\caption{Cyclic stress-softening with residual strain and creep of residual strain.}
\label{fig:12y}
\end{figure}

\begin{figure}[ht]
\centering
\includegraphics[width=18cm,height=10cm]{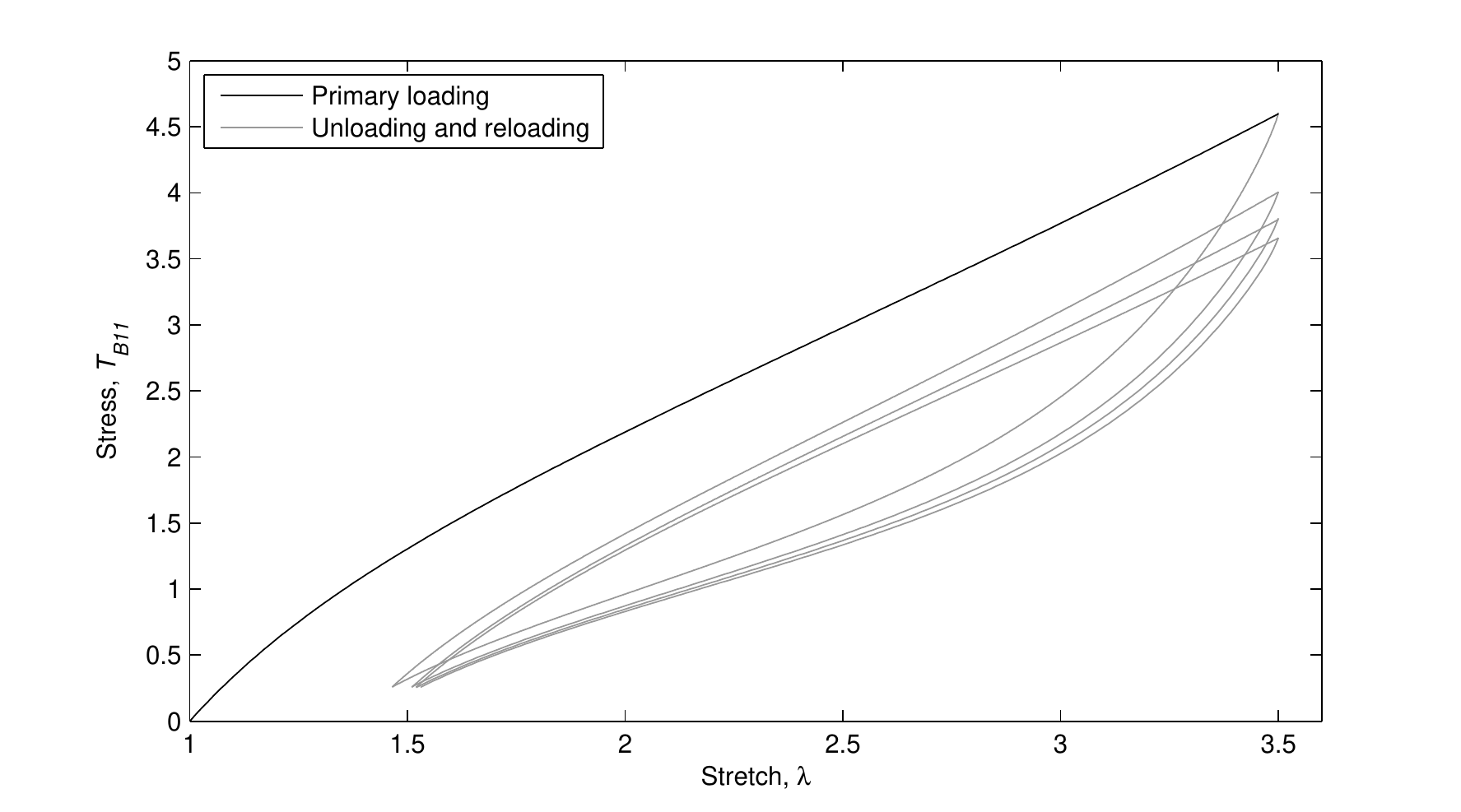}
\vspace{-20pt}
\caption{Cyclic stress-softening for a non-zero stress.}
\label{fig:12yy}
\end{figure}

 Figures \ref{fig:10y}, \ref{fig:11y}, \ref{fig:12y} and \ref{fig:12yy} depict cyclic stress-softening paths in uniaxial tension for a variety of typical materials.  They have been obtained by employing variations of the following constants and functions, which have realistic values but do not correspond to any known materials:
\[
N=27.9, \quad \mu=1.2, \quad r_1 = r_2=2,
\]
\[
{A}_0=-0.115 \quad a(t)=t, \quad a_1=0.4,
\]
\[
A_{1,2}(t)=\left\{ \begin{array}{clrr}
-0.01\log(0.2t)& \\
-0.01\log(0.55t)& \\
\end{array}\right.
d_\omega=\left\{ \begin{array}{clrr}
-0.077\\
-0.035\\
\end{array}\right.
\mu b_\omega=\left\{ \begin{array}{clrr}
2.38\\
50.0\\
\end{array}\right.
\vartheta_\omega=\left\{\begin{array}{clrr}
0.800& \textrm{unloading},\\
1.000& \textrm{reloading}.\\
\end{array}\right.
\]

 In Figure \ref{fig:10y} the Mullins effect is depicted with stress-relaxation but no residual strain or creep and so equation (\ref{eq:16z}) is used for the stress.  In Figure \ref{fig:11y} we depict the Mullins effect with residual strain and the creep causing it and so use equation  (\ref{eq:22z})  for the stress.  Figure \ref{fig:12y} represents the Mullins effect in a material displaying   stress-softening, hysteresis, stress relaxation,  residual strain and creep of residual strain.  This is the full model derived here and so equation  (\ref{eq:27z}) is used  for the stress. \tcr{In Figure  \ref{fig:12yy}  we depict the cyclic stress softening of a carbon filled rubber vulcanizate in which the unloading proceeds to a positive stress level rather than all the way to zero.  The model copes well, using the same value of $W_\M$, with this  situation, which may arise in an engineering application where a rubber component, such as a spring or damper, is subject to a cyclic uniaxial tension with a continuous positive strain being applied.}

\newpage
\subsection{Comparison with  experimental data}
\label{sec:experimentaldata}
\begin{figure}
\centering
\includegraphics[width=18cm,height=10cm]{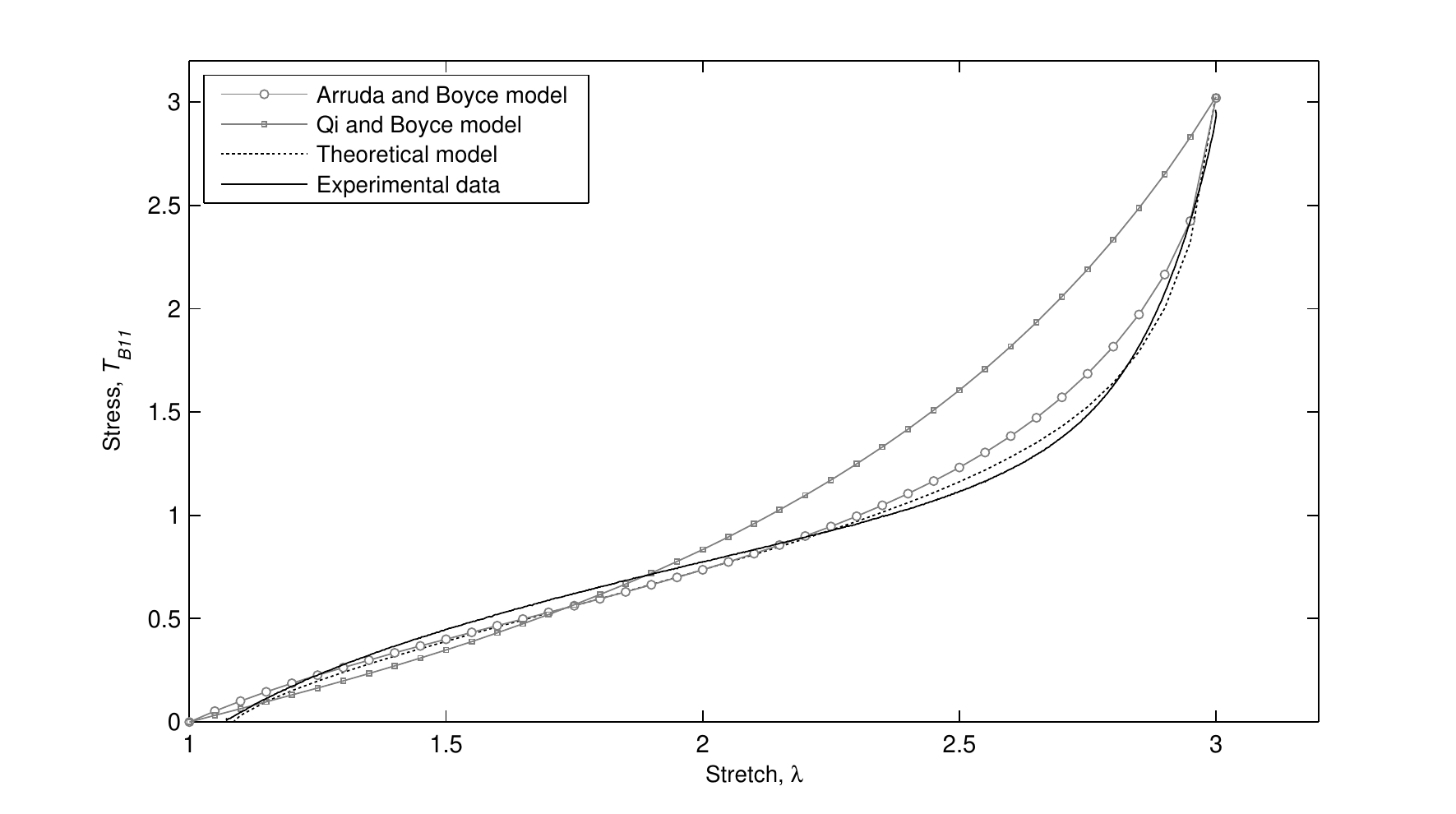}
\vspace{-20pt}
\caption{Comparison of the first unloading path with other constitutive models.}
\label{fig:13y}
\end{figure}

In Figure \ref{fig:13y} we compare the model developed here with the \cite{arruda} model  and the model of  \cite{qi}. We are  comparing only the first unloading path as we have been unable to identify any other model in the literature that takes into account either the first reloading path, path $C$ of Figure \ref{fig:4y}, or further cyclic loading, see also Figure \ref{fig:4y}. In the models that have appeared in the literature the unloading/reloading paths for the cyclic process are treated as being one and the same path.    We have not compared the primary loading paths as our constitutive equation, namely equation (\ref{eq:6z}), for this path is the same  as that presented by \cite{arruda} and comparisons with this model have been well documented, see \cite{zuniga}.

\begin{figure}
\centering
\includegraphics[width=18cm,height=10cm]{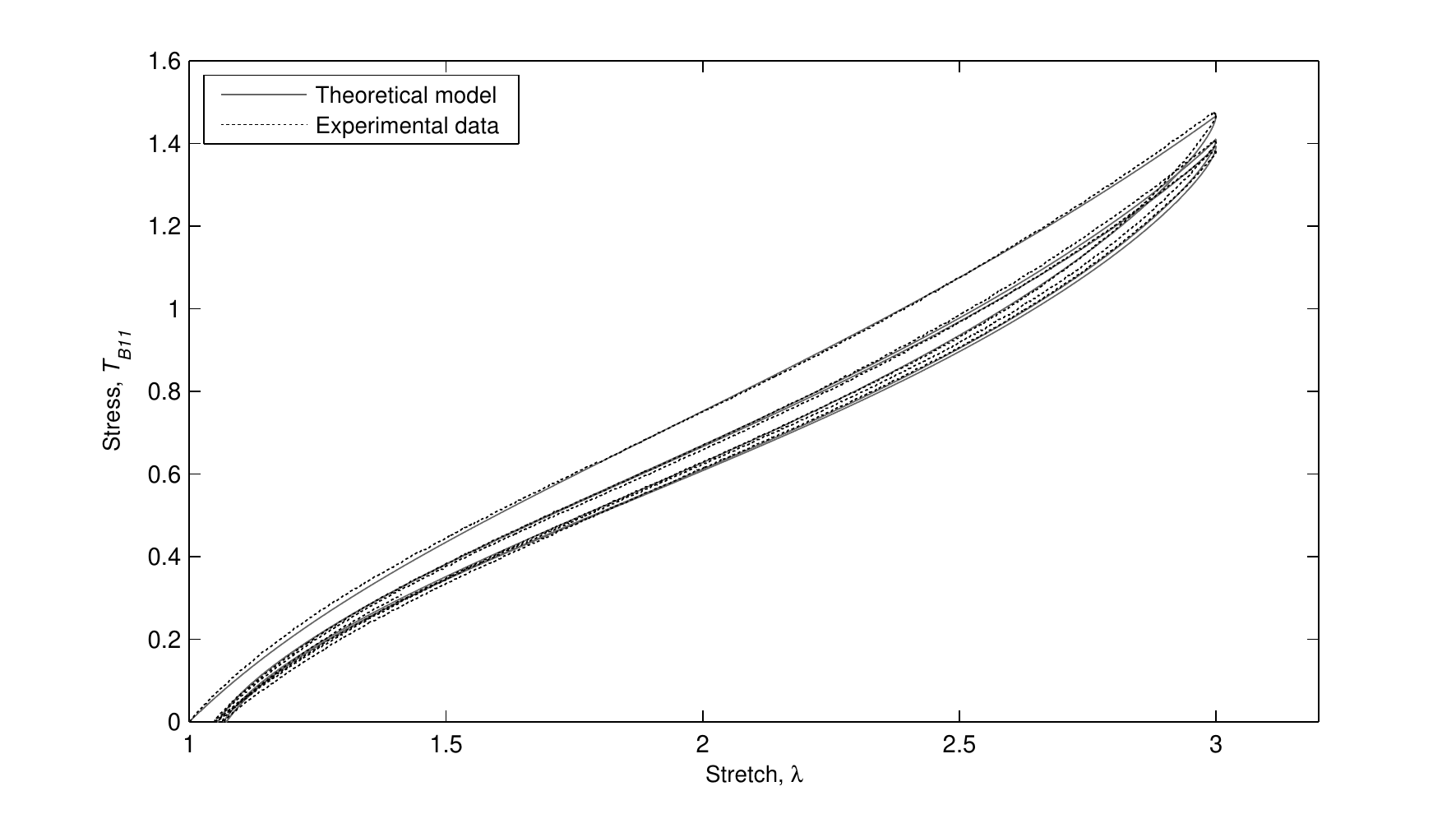}
\vspace{-20pt}
\caption{Comparison of our theoretical model with experimental data of \cite{dorfmann}, particle-reinforced
specimen with 1 phr of carbon black.}
\label{fig:14y}
\end{figure}

\begin{figure}
\centering
\includegraphics[width=18cm,height=10cm]{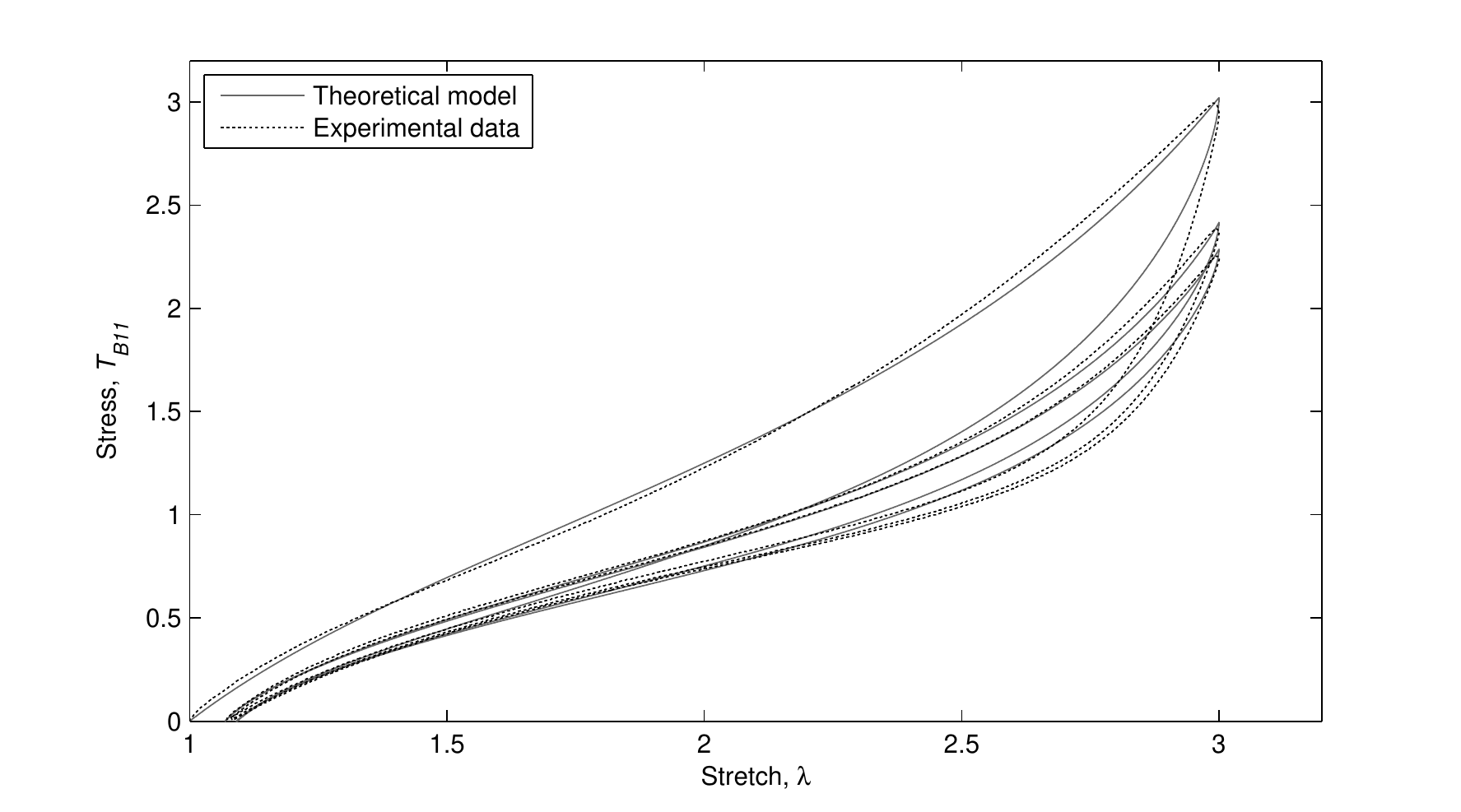}
\vspace{-20pt}
\caption{Comparison of our theoretical model  with experimental data of \cite{dorfmann}, particle-reinforced
specimen with 20 phr of carbon black.}
\label{fig:15y}
\end{figure}

\begin{figure}
\centering
\includegraphics[width=18cm,height=10cm]{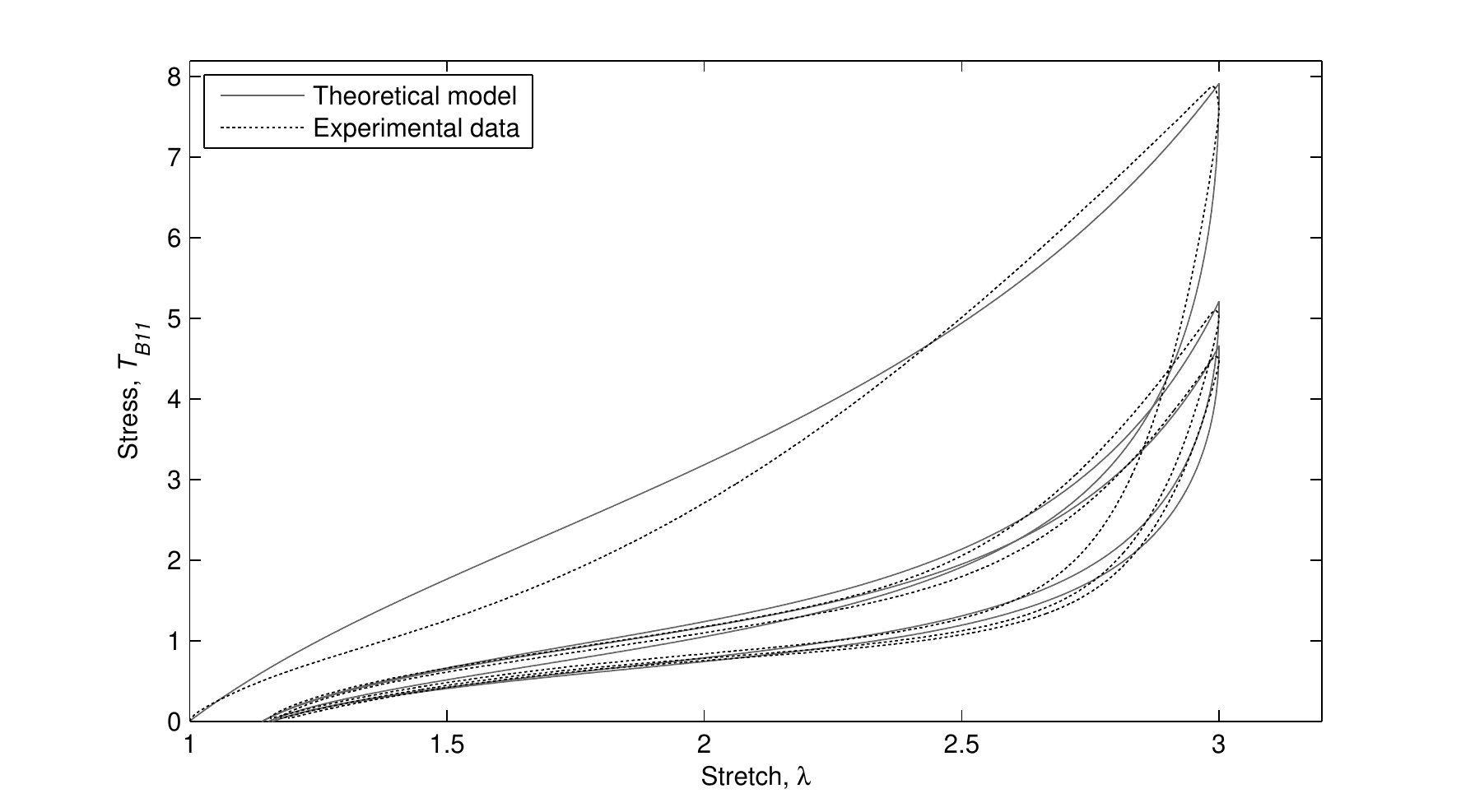}
\vspace{-20pt}
\caption{Comparison of our theoretical model  with experimental data of \cite{dorfmann}, particle-reinforced
specimen with 60 phr of carbon black.}
\label{fig:16y}
\end{figure}

We see from Figure \ref{fig:13y} that our model provides a slight improvement compared to the model of  \cite{arruda}, though it should be emphasized that our model is for cyclic stress-softening.   In Figures \ref{fig:14y}, \ref{fig:15y} and \ref{fig:16y} we provide a comparison of our  constitutive model with experimental data for various concentrations of carbon-black filled natural rubber vulcanizates.   This data was kindly provided by Professor A. L. Dorfmann  and was presented previously by  \cite{dorfmann}.

Figure \ref{fig:14y} was obtained by using the following constants and functions in our model:
\[
N=9.3, \quad \mu=0.375, \quad r_1=r_2=1.47,
\]
\[
{A}_0=-0.001, \quad a(t)=t, \quad a_2=0.4,
\]
\[
A_{1,2}(t)= \left\{ \begin{array}{clrr}
-0.002\log(0.7t)\\
-0.002\log(1.4t)\\
\end{array}\right.
d_\omega=\left\{ \begin{array}{clrr}
-0.0005\\
-0.0003\\
\end{array}\right.
\mu b_\omega=\left\{ \begin{array}{clrr}
28.0\\
28.0\\
\end{array}\right.
\vartheta_\omega=\left\{\begin{array}{clrr}
0.55& \textrm{unloading},\\
0.80& \textrm{reloading}.\\
\end{array}\right.
\]

Figure \ref{fig:15y} was obtained by using the following constants and functions in our model:
\[
N=5.7, \quad \mu=0.56, \quad r_1=r_2=1.31,
\]
\[
{A}_0=-0.0017, \quad a(t)=t, \quad a_2=1.4,
\]
\[
A_{1,2}(t)= \left\{ \begin{array}{clrr}
-0.0115\log(0.7t)\\
-0.0115\log(1.4t)\\
\end{array}\right.
d_\omega=\left\{ \begin{array}{clrr}
-0.0014\\
-0.0007\\
\end{array}\right.
\mu b_\omega=\left\{ \begin{array}{clrr}
10.5\\
10.5\\
\end{array}\right.
\vartheta_\omega=\left\{\begin{array}{clrr}
0.55& \textrm{unloading},\\
0.85& \textrm{reloading}.\\
\end{array}\right.
\]

Figure \ref{fig:16y} was obtained by using the following constants and functions in our model:
\[
N=5.5, \quad \mu=1.41, \quad r_1=r_2=1.11,
\]
\[
{A}_0=-0.055, \quad a(t)=t, \quad a_2=0.4,
\]
\[
A_{1,2}(t)= \left\{ \begin{array}{clrr}
-0.06\log(0.7t)\\
-0.06\log(1.4t)\\
\end{array}\right.
d_\omega=\left\{ \begin{array}{clrr}
-0.012\\
-0.010\\
\end{array}\right.
\mu b_\omega=\left\{ \begin{array}{clrr}
1.8\\
3.2\\
\end{array}\right.
\vartheta_\omega=\left\{\begin{array}{clrr}
0.55& \textrm{unloading},\\
0.95& \textrm{reloading}.\\
\end{array}\right.
\]

Figure \ref{fig:14y} represents a specimen of rubber with a very low filler concentration, only 1phr, and the model we have  developed  provides an extremely accurate representation of the experimental data of \cite{dorfmann}. 
\tcr{We observe in Figure \ref{fig:14y}  that the unloading and reloading paths are almost parallel. This is directly reflected in the choice of parameters used within the model,  there being only a small variation in the creep parameter $d_\omega$, see equation (\ref{eq:26z}),  used in unloading and reloading.   There is a larger variation in the parameter $\vartheta_\omega$, see equation (\ref{eq:10z}), because this is  largely the parameter that governs the degree of hysteresis between the unloading and reloading paths.}  

\tcr{The experimental data presented in Figure \ref{fig:15y} is for a carbon filled rubber vulcanizate with an increased concentration of 20 phr but again the unloading and reloading paths are almost parallel. This is once more reflected in the choice of parameters used within the model with only $\vartheta_\omega$ varying significantly between unloading and reloading.} 

\tcr{In Figure \ref{fig:16y} we present experimental data for a filled rubber vulcanizate with the much higher concentration of 60 phr of carbon black and observe that now the unloading and reloading paths are no longer parallel. This loss of symmetry is seen in the parameters used within the model as now $d_\omega$ and  $\mu b_\omega$  vary between unloading and reloading, as well as $\vartheta_\omega$.}

From Figures \ref{fig:15y} and \ref{fig:16y} we see that as the filler concentration increases the \cite{arruda} model overestimates the stress on the primary loading path.   We also observe that the accuracy of the unloading and reloading paths decreases with increased filler concentration.

\section{Conclusion}
\label{sec:conclusion}

The model presented here appears to be the first in which a stress-softening and residual strain model has been combined with the Arruda-Boyce eight-chain model of elasticity in order to develop a model that is capable of representing the Mullins effect for an isotropic, incompressible, hyperelastic material.    Figures \ref{fig:10y}, \ref{fig:11y} and \ref{fig:12y} show that the model has been quite successful.

 We have considered alternative approaches to modelling the residual strain but they do not appear able to  replicate the creep associated with the Mullins effect to the same degree of accuracy as the model presented here.   For example, if the exponent in equation (\ref{eq:17z}), and hence in the last line of equation (\ref{eq:29z}), differs much from $-1$ then it is not possible to model the creep at all well.  This argues well for the model of \cite{bergstrom}.

From Figure \ref{fig:12y}, we see that the model developed here provides a good representation of the Mullins effect for uniaxial tension of an isotropic rubber-like material. This model has been developed in such a way that any of the salient inelastic features can be excluded and the integrity of the model  still be maintained.  

We remark that Figures \ref{fig:15y} and \ref{fig:16y} show limited agreement with experiment though modelling correctly the broad features.  This may be due to the fact that after an applied uniaxial tension the material is effectively  transversely isotropic, rather than purely isotropic, because of bond-breaking and realignment.   The direction of uniaxial tension would therefore become the preferred direction of transverse isotropy.   We hope to present in a future paper the  extension of the present model to transversely isotropic materials.   

 \cite{park} and other authors have noted that unfilled vulcanized natural rubber shows negligible anisotropy as is consistent with Figure \ref{fig:14y}, which shows excellent agreement with experimental data for a very low concentration of carbon black. 

The results presented here are capable of  extension to  equibiaxial tension and pure shear for multi-cyclic stress-strain  loading.  We hope to discuss these matters  in a future paper.

\section*{Acknowledgements}
One of us (SRR) is grateful to  the University of East Anglia for the award of a PhD studentship. The authors thank  Professor Luis Dorfmann for most kindly supplying experimental data. \tcr{Furthermore, we would like to thank the reviewers for their constructive comments and suggestions}.

\bibliographystyle{model2-names}
\bibliography{BIBLIOGRAPHYa}

\end{document}